%% file: mnras_template.tex
\documentclass[fleqn,usenatbib]{mnras}

\usepackage{newtxtext,newtxmath}

\usepackage[T1]{fontenc}

\DeclareRobustCommand{\VAN}[3]{#2}
\let\VANthebibliography\thebibliography
\def\thebibliography{\DeclareRobustCommand{\VAN}[3]{##3}\VANthebibliography}

\usepackage{graphicx}	
\usepackage{amsmath}	
\usepackage{CJKutf8}    
\usepackage{subfig}     
\usepackage{pdflscape}  
\usepackage{afterpage}

\newcommand{\ujy}{$\mu$Jy\,beam$^{-1}$}
\newcommand{\mjy}{mJy\,beam$^{-1}$}
\newcommand{\chinese}[1]{\begin{CJK}{UTF8}{gbsn}#1\end{CJK}}
\newcommand{\replace}[1]{{#1}}

\title[ASKAP Transients on Minute Timescale]{Radio Variable and Transient Sources on Minute Timescales in the ASKAP Pilot Surveys}

\author[Y. Wang et al.]{Yuanming Wang (\chinese{王远明}),$^{1,2,3}$\thanks{E-mail: ywan3191@uni.sydney.edu.au}
Tara Murphy,$^{1,3}$\thanks{E-mail: tara.murphy@sydney.edu.au}
Emil Lenc,$^{2}$
Louis Mercorelli,$^{4}$ 
Laura Driessen,$^{1,5}$
\newauthor
Joshua Pritchard,$^{1,2,3}$
Baoqiang Lao (\chinese{劳保强}),$^{6,7}$
David L.\ Kaplan,$^{8}$
Tao An (\chinese{安涛}),$^{7}$
Keith W.\ Bannister,$^{2}$
\newauthor
George Heald,$^{5}$
Shuoying Lu (\chinese{卢铄迎}),$^{9}$ 
Artem Tuntsov,$^{10}$
Mark Walker,$^{10}$
and Andrew Zic$^{2}$
\\
$^{1}$Sydney Institute for Astronomy, School of Physics, University of Sydney, Sydney, NSW 2006, Australia\\
$^{2}$Australia Telescope National Facility, CSIRO, Space and Astronomy, PO Box 76, Epping, NSW 1710, Australia\\
$^{3}$ARC Centre of Excellence for Gravitational Wave Discovery (OzGrav), Hawthorn, Victoria, Australia\\
$^{4}$Sydney Informatics Hub, The University of Sydney, NSW 2006, Australia\\
$^{5}$Australia Telescope National Facility, CSIRO, Space and Astronomy, PO Box 1130, Bentley WA 6102, Australia\\
$^{6}$School of Physics and Astronomy, Yunnan University, Kunming, 650091, China \\
$^{7}$Shanghai Astronomical Observatory, Key Laboratory of Radio Astronomy, Chinese Academy of Sciences, Shanghai, China\\
$^{8}$Center for Gravitation, Cosmology, and Astrophysics, Department of Physics, University of Wisconsin-Milwaukee, P.O. Box 413, Milwaukee, WI 53201, USA\\
$^{9}$School of Information and Communication, Guilin University of Electronic Technology, Guilin 541004, China\\
$^{10}$Manly Astrophysics, 15/41-42 East Esplanade, Manly 2095, Australia
}

\date{Accepted XXX. Received YYY; in original form ZZZ}

\pubyear{2015}

\begin{document}
\label{firstpage}
\pagerange{\pageref{firstpage}--\pageref{lastpage}}
\maketitle

\begin{abstract}
We present results from a radio survey for variable and transient sources on 15-min timescales, using the Australian SKA Pathfinder (ASKAP) pilot surveys. 
The pilot surveys consist of 505\,h of observations conducted at around 1\,GHz observing frequency, with a total sky coverage of 1,476\,deg$^2$. 
Each observation was tracked for approximately 8--10\,h, with a typical rms sensitivity of $\sim$30\,\ujy\ and an angular resolution of $\sim$12\,arcsec. 
The variability search was conducted within each 8--10\,h observation on a 15-min timescale. 
We detected 38 variable and transient sources. 
Seven of them are known pulsars, including an eclipsing millisecond pulsar, PSR~J2039$-$5617. 
Another eight sources are stars, only one of which has been previously identified as a radio star. 
For the remaining 23 objects, 22 are associated with active galactic nuclei or galaxies (including the five intra-hour variables that have been reported previously), and their variations are caused by discrete, local plasma screens. 
The remaining source has no multi-wavelength counterparts and is therefore yet to be identified. 
This is the first large-scale radio survey for variables and transient sources on minute timescales at a sub-mJy sensitivity level. 
We expect to discover $\sim$1 highly variable source per day using the same technique on the full ASKAP surveys. 

\end{abstract}




\section{Introduction}

Radio variable and transient sources usually imply extreme astrophysical environments including (intrinsic) strong magnetic activities, explosions, accretion, and (extrinsic) propagation effects caused by inhomogeneities in the ionised interstellar medium (ISM) (e.g., see \citealt{Cordes2004,Murphy2013}). 
Early untargeted surveys for radio variables have led to a few discoveries \citep[e.g.,][]{Levinson2002,Hyman2005,Gal-Yam2006}. 
However, these surveys are historically limited by narrow fields-of-view (FoV), poor sensitivity, or sparse observing cadence \citep[e.g.,][]{Bower2007,Bannister2011,Bell2011}. 
Recently-built or upgraded telescopes have significant improvements, allowing for large-scale radio transient surveys with greatly increased sensitivity, such as
the Variables and Slow Transients (VAST) survey using the Australian SKA Pathfinder (ASKAP; \citealt{Hotan2021}) at 0.7--1.8\,GHz \citep{Murphy2013,Murphy2021}; 
the ThunderKAT image-plane transients programme using MeerKAT at $\sim$GHz \citep{Fender2016}; 
the Murchison Widefield Array Transients Survey at 154\,MHz \citep{Bell2019}; 
and the Amsterdam-ASTRON Radio Transients Facility and Analysis Center (AARTFAAC) all-sky monitor using the Low Frequency Array (LOFAR) at 10--90\,MHz \citep{Prasad2016}. 
Other large-scale radio continuum surveys such as 
the TIFR Giant Metrewave Radio Telescope Sky Survey \citep[TGSS;][]{Intema2017}, 
the LOFAR Two-meter Sky Survey \citep[LoTSS][]{Shimwell2017}, 
the Very Large Array (VLA) Sky Survey \citep[VLASS;][]{Lacy2020}, 
the GaLactic and Extragalactic All-sky Murchison Widefield Array (MWA) survey eXtended \citep[GLEAM-X;][]{Hurley-Walker2022a} can also be used for transients searches. 
These new-generation telescopes give great opportunities to explore the parameter space that was poorly-explored before, e.g., short-timescale radio transients ($\lesssim$hours).  
Compared to previous surveys which usually performed on timescales $\gtrsim$1 day, transient and variable sources on shorter timescales $\sim$minutes would be considerably different, mainly including flaring stars to pulsars, or external effects such as enhanced scintillation. 

A wide range of stellar objects can emit highly circularly polarised radio flares on timescales of $\sim$minutes (see a review from \citealt{Gudel2002}). 
These flares are most likely generated from coherent emission process, e.g. plasma emission or electron cyclotron maser (ECM) emission, allowing constraints on the stellar magnetosphere through measured electron density (plasma emission) or magnetic field strength (ECM emission; see reviews from \citealt{Dulk1985} and \citealt{Osten2008}).
A recent ASKAP large-scale untargeted survey identified more than 30 radio flaring stars, and 23 of them have no previous radio detections \citep{Pritchard2021}. 
These stars have a broad range from M-dwarfs to magnetically chemically peculiar B-type stars, showing that wide-field surveys are efficient to discover previously unknown radio flaring stellar objects.

Pulsars are rapidly rotating neutron stars with beamed radio (and/or high-energy) emission from their magnetic poles (periods from milliseconds to seconds).
The discovery of pulsars opened a new research field that allow to study neutron star physics, general relativity, and the interstellar medium \citep[see][]{Lorimer2012}. 
Their radio emission is thought to be coherent and the emission mechanisms are not fully understood yet (see a review from \citealt{Philippov2022}). 
The variability of pulsars on $\sim$minutes timescales can be from a range of different origins, including deep nulling (pulse energy suddenly drops to zero and then suddenly returns back to its normal state; e.g., \citealt{Backer1970}), eclipses in a binary system (e.g., \citealt{Camilo2000}), and diffractive scintillation \citep[e.g.,][]{Narayan1992}. 
It is possible to discover new pulsars from radio continuum surveys based on their variability, polarisation, and multi-wavelength properties \citep[e.g.,][]{Kaplan2019,Wang2022,Sobey2022}. 

Intrahour variables (IHVs) are extreme scintillating active galactic nuclei (AGN) with large amplitude modulations ($\gtrsim$10\%) and short timescales ($\lesssim$ hours), caused by small-scale irregularities in the ionised ISM with required pressure fluctuations much higher than those in the typical diffuse ISM \citep[e.g.,][]{Kedziora-Chudczer1997,Dennett-Thorpe2002,Walker2017,Bignall2019,Oosterloo2020}. 
There are studies showing that these high-pressure clouds are close to the Earth (a few parsecs) and their physical nature is still unclear \citep[e.g.,][]{Dennett-Thorpe2003,Wang2021b}.
IHV provides an opportunity to explore the possible origin of discrete plasma in the solar neighbourhood, as well as small-scale AGN strucutre. 

More recently, there have been discoveries of ultra-long period neutron stars \citep[e.g.,][]{Hurley-Walker2022b,Caleb2022}. 
Their periods are unusually slow (18.18\,min and 76\,s) located beyond the `death line' as defined by the inner vacuum-gap curvature radiation models for radio emission \citep{Chen1993}, challenging our understanding of how these systems evolve. 
These discoveries highlighted the possibility of identifying new populations on these poorly-explored short timescales. 

Only a few radio transient surveys have covered the timescales of $\sim$minutes and resulted in a few (or no) convincing transients\footnote{See a summary table in \citet{Mooley2016} and \url{http://www.tauceti.caltech.edu/kunal/radio-transient-surveys/index.html}}. 
Early surveys on minute-timescales were mainly at low frequencies (tens of MHz) with low angular resolution ($\sim$degs) and poor sensitivity \citep[e.g.,][]{Lazio2010,Obenberger2014}. 
The first survey explored minutes-timescales at GHz is by \citet{Thyagarajan2011}. 
They conducted an analysis for variable and transient radio sources down to $\sim$mJy levels over timescales from a few minutes to years using overlapped adjacent snapshot images ($\sim$3-min interval) of the Faint Images of the Radio Sky at Twenty-cm (FIRST) survey \citep{Becker1995} at 1.4\,GHz. 
They found several stars and pulsars accompanied with a large sample of unclassified or unidentified objects due to lack of multi-wavelength identification. 
\citet{Stewart2016} searched for low-frequency transients on timescales of 0.5--297\,min using LOFAR at a sensitivity of $\sim$Jy level. 
They found one transient with a duration of a few minutes and a flux density of 15--25\,Jy at 60\,MHz.
The nature of this transient is still unclear. 
\citet{Hobbs2016} ran transient detection on 2-min snapshot images from 13\,h ASKAP-BETA observation for an intermittent pulsar PSR~J1107$-$5907. 
They detected the target pulsar but no other transient events. 
\citet{Chiti2016} performed a search for radio transients on timescales from seconds to minutes using more than 200\,h VLA archival data at 5 and 8.4\,GHz in the Galactic centre, with a typical detection threshold of $\sim$100\,mJy. 
They detected two promising, unclassified candidates and suggested them as either a new astrophysical source or a subtle imaging artefact. 

The key to achieve a transient survey on short-timescales is a telescope with high instrumental sensitivity, large FoV, and more importantly, great instantaneous ($u,v$) coverage to build a good-quality sky model on a short sampling time. 
ASKAP is a survey instrument designed using phased array feed (PAF) technology \citep{DeBoer2009,Hotan2021}, which provides 36 beams for each of 36 12-m antennas to achieve a wide FoV and good sensitivity. 
The array configuration provides excellent instantaneous ($u,v$) coverage\footnote{\url{https://www.atnf.csiro.au/projects/askap/newdocs/configs-3.pdf}}, making snapshot imaging surveys (e.g. at 15-min time-scale) possible. 
With the ASKAP telescope we are now able to discover short-timescale transients (seconds to hours) in a more systematic way. 
The improved large-scale multiwavelength surveys can also provide valuable information to identify detected transients. 

In this paper we describe the first large-scale survey for radio variable and transient sources on a timescale of 15\,min using ASKAP pilot survey data. 
The data we used are from a large set of pilot survey projects, total of 505\,h covering a total of 1,476\,deg$^2$ sky area at a sub-mJy sensitivity. 
In Section~\ref{sec:observations_processing} we describe the survey data and processing process, with the following variability search described in Section~\ref{sec:variability_search}. 
In Section~\ref{sec:results} we present results from the survey and summarise properties of our candidates, and in Section~\ref{sec:discussion} we discuss our sources, the variability rate, and the application of this approach to future surveys.

\section{Observations and data processing}
\label{sec:observations_processing}

ASKAP began its pilot survey program on 2019~July~15 \citep[see][Section~15]{Hotan2021}. 
This involved 11 continuum and spectral-line surveys, resulting in more than 1,000 hours of observations. 
The pilot surveys used their requested configurations and parameters, providing a rich archival database\footnote{All of the ASKAP data can be accessed through
the CSIRO ASKAP Science Data Archive (CASDA) \url{https://research.csiro.au/casda/}} suitable for various science goals. 

\subsection{Observations}

\begin{figure*}
    \includegraphics[width=0.8\textwidth]{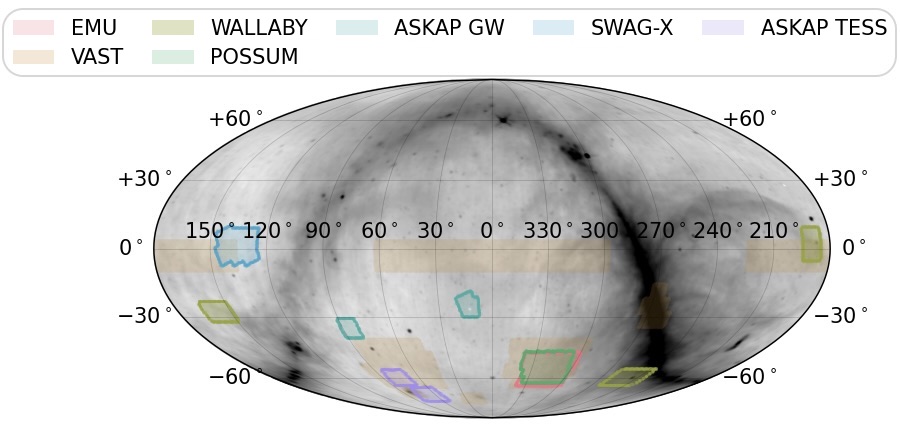}
    \caption{Sky coverage of ASKAP pilot survey data we used in this study. The sky map is plotted with J2000 equatorial coordinates in the Mollweide projection and the background diffuse Galactic emission at 887.5\,MHz is modelled from \citet{Zheng2017}. }
    \label{fig:sky_map.png}
\end{figure*}

\begin{figure*}
    \includegraphics[width=0.9\textwidth]{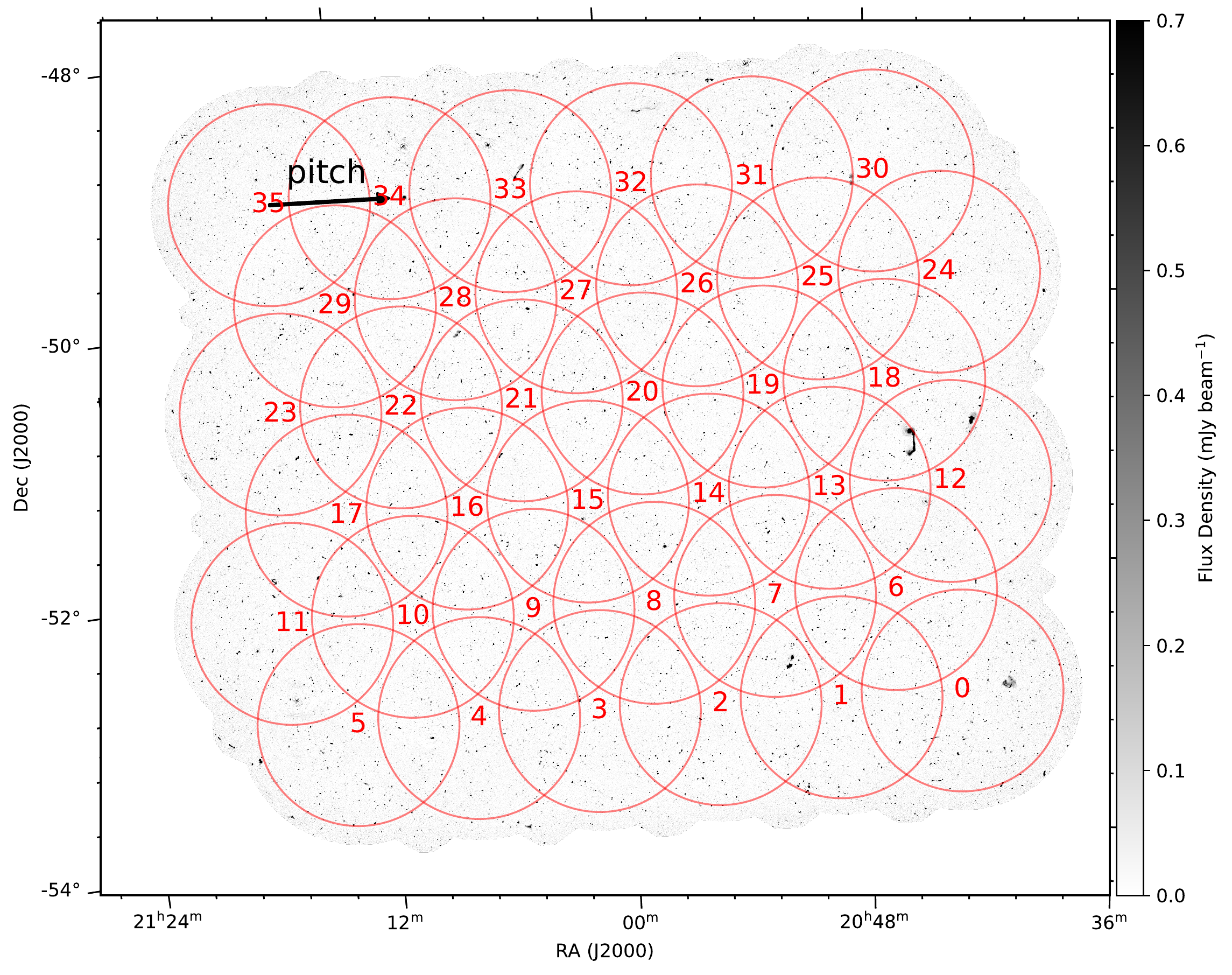}
    \caption{An example ASKAP image from one of the EMU observations (SB9287). The red circles represent the arrangement of the 36 beams in the \texttt{closepack36} configuration, numbered from 0 to 35. The pitch spacing is 0.9\degr\ and the actual full-width half maximum for the primary beam is approximately 1.5\degr\ at 943.5\,MHz (as represented by the red circle). The background image is a mosaiced version made by combining the separate images for each beam. Note that we processed and analysed each beam independently across the whole paper. }
    \label{fig:EMU_example}
\end{figure*}

We used ASKAP pilot survey data. 
Our aim was to search for intra-observation transients on 15-min time-scales in the image plane. 
We selected long observations ($>$1\,h) that used the standard correlator mode (i.e., not the zoom mode) and no interleaving. 
Figure~\ref{fig:sky_map.png} shows the total sky coverage of selected survey data. 
We mainly used observations from the following surveys: 

\begin{description}
    
    \item[\textbf{EMU}] The Evolutionary Map of the Universe (EMU; project code AS101) aims to make a deep radio continuum all-sky map \replace{from declination of $-90\degr$ to $+30\degr$} and expects to detect and catalog about 70\,million galaxies \citep{Norris2011}. 
    The EMU pilot survey was observed at a central frequency of 943.5\,MHz with a bandwidth of 288\,MHz. 
    The integration time for each field is 10\,h, reaching a sensitivity of 25--30\,\ujy at a spatial resolution of 11--18\,arcsec \citep{Norris2021}. 
    The sky area of the EMU pilot survey \replace{was within} the Dark Energy Survey (DES; \citealt{Abbott2018}) field, which is helpful to identify any optical counterpart for a detected radio source. 
    
    \item[\textbf{WALLABY}] The Widefield ASKAP L-band Legacy All-sky Blind surveY (WALLABY; project code AS102) is a next-generation survey of neutral hydrogen (HI) in the Local Universe \citep{Koribalski2020}. 
    The ASKAP pilot survey for WALLABY was conducted at a central frequency of 1367.5\,MHz with a bandwidth of 144\,MHz, targeting the Hydra cluster, the NGC 4636 group, and the Norma cluster.
    The integration time for each observation was 8\,h. 
    WALLABY used a different beam configuration, \texttt{square\_6x6} (a square grid arrangement), compared to other surveys such as EMU. 
    The latter normally used \texttt{closepack36} arrangement (see Figure~\ref{fig:EMU_example}), which offsets alternating rows for a more uniform sensitivity (see Section~9 in \citealt{Hotan2021}). 
    
    \item[\textbf{POSSUM}] The Polarisation Sky Survey of the Universe's Magnetism (POSSUM; project code AS103) aims to measure the Faraday rotation of three million extragalactic radio sources over 30,000 square degrees, improving the understanding of astrophysical magnetism \citep{Gaensler2010}. 
    The ASKAP pilot survey for POSSUM targeted to same sky region with EMU, tracked for 8\,h but at a higher central frequency of 1367.5\,MHz with a bandwidth of 144\,MHz. 
    
    \item[\textbf{ASKAP GW}] The ASKAP Follow-up Observations for Gravitational Wave Counterparts (ASKAP GW; project code AS111) aimed to conduct searches for radio afterglows of gravitational wave events GW190814 \citep{Dobie2019,Dobie2022}. 
    It contains 10 observations separated by days to months. 
    The integration time for each observation is approximately 10.5\,h, achieving a sensitivity of 35--40\,\ujy at a spatial resolution of $12\times10$\,arcsec \citep{Dobie2022}. 
    The first four epochs (SB9602, 9649, 9910, and 10463) were used by \citet{Wang2021b} to demonstrate intra-observation transients searches with ASKAP, resulting in the discovery of a long, thin, and straight plasma filament revealed by five IHVs in a line on the sky. 
    
    \item[\textbf{SWAG-X}] The GAMA-09 + X-ray survey (SWAG-X; project code AS112) was designed to cover the GAMA\footnote{\url{http://www.gama-survey.org/}} and eROSITA\footnote{\url{https://www.mpe.mpg.de/eROSITA}} Final Equatorial-Depth Survey (Moss et al., in prep). 
    SWAG-X tiles were located farther north than most other data, and were observed at a central frequency of 887.5\,MHz with a bandwidth of 288\,MHz. 
    
    \item[\textbf{Other ToOs}] We also included two observations from target of opportunity observations (ToOs) or guest observations (project code AS113). 
    The two ASKAP fields were scheduled to align with Sector~36 observed by the \textit{Transiting Exoplanet Survey Satellite} (\textit{TESS}; \citealt{Ricker2015}). 
    The integration time for each field is 13\,h, achieving a rms sensitivity of 21--24\,\ujy at a central frequency of 887.5\,MHz (see details in \citealt{Rigney2022}). 

\end{description}

Table~\ref{tab:observations} lists details for 52 observations selected from the above surveys. 
Each observation was approximately 8--10\,h, with a FoV up to 66\,deg$^2$ at the low-frequency band (887.5/943.5\,MHz) and down to 29\,deg$^2$ at high-frequency band (1367.5\,MHz). 
An example image (from one of EMU observations) is shown in Figure~\ref{fig:EMU_example}, overlapped with a typical beam configuration \texttt{closepack36} in 0.9\degr\ pitch\footnote{Pitch represents the spacing of neighbouring beam centre. See Figure~\ref{fig:EMU_example}}. 
The total observing time was 505\,h and the overall (unique) sky coverage reached 1,476\,deg$^2$ (note POSSUM was overlapped with EMU), including an extremely deep field S190814bv with total tracking time of approximately 115\,h.

\input{table_observations}

\subsection{Processing workflow}
\label{subsec:workflow}

\begin{figure*}
    \includegraphics[width=\textwidth]{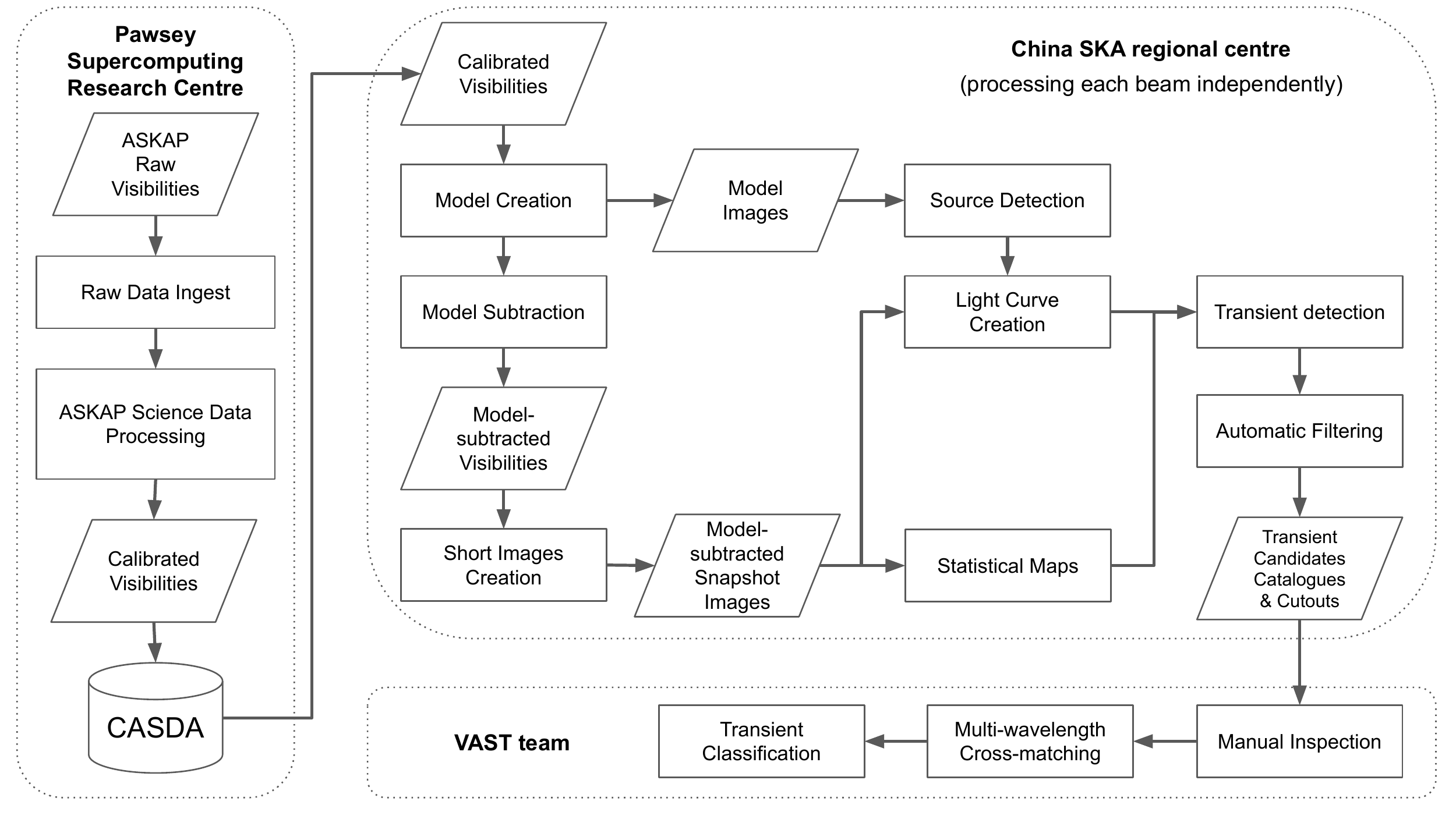}
    \caption{Illustration of the data processing workflow. A description of each stage is given in Section~\ref{subsec:workflow}. }
    \label{fig:flowchart}
\end{figure*}

Figure~\ref{fig:flowchart} shows the overall data processing workflow, from raw visibilities, standard ASKAP outputs, to our final products. 
The processing consists of three parts performed at different sites: 
the standard ASKAP processing at the Pawsey Supercomputing Research Centre\footnote{\url{https://pawsey.org.au/}} in Western Australia, 
our dedicated processing for transients search at the China SKA regional centre, 
and the final candidates inspection and investigation by VAST team on the local machine. 
Below we give a description of each processing step.  

\begin{description}
    \item[\textbf{Raw data ingest}]
    Raw visibility data from the ASKAP correlator hardware are automatically ingested to the Pawsey Supercomputing Research Centre. 
    Each of the PAF beams were split into its own measurement set, i.e., total of 36 measurement sets for one observation. 
    
    \item[\textbf{ASKAP science data processing}] 
    Each ASKAP observation was processed using the \textsc{askapsoft} pipeline \citep{Whiting2020,Guzman2019} following the standard procedure described in \citet[][Section~11]{Hotan2021}. 
    This includes bandpass calibration (using a separate observation of PKS~B1934$-$638), radio-frequency interference (RFI) flagging, initial imaging and self-calibration, wide-field imaging (per beam), and linear mosaicking (to create a full-field image). 
    After the mosaicked full-field images have been produced, source finding is run to create catalogues using \textsc{selavy} \citep{Whiting2012}. 

    \item[\textbf{Deposit to CASDA}] 
    The standard ASKAP data products, including calibrated visibilities at continuum resolution, mosaicked full-field images, and corresponding \textsc{selavy} catalogues, are deposited to CSIRO ASKAP Science Data Archive (CASDA\footnote{\url{https://research.csiro.au/casda/}}).
\end{description}

After the data have been uploaded to CASDA, we download the calibrated visibilities through the CASDA virtual observatory (VO) services\footnote{\url{https://casda.csiro.au/casda_vo_tools/tap}} to the China SKA regional centre prototype (CSRC-P; \citealt{An2019,An2022}) in Shanghai. 
The data transferring takes about 2--3\,h for a $10$\,h observation. 
We then conduct our dedicated processing for transients search on 15-min timescale. 
The CSRC-P is equipped with 23 multi-core x86 CPU nodes with memory size of 0.77--1\,TB each, allowing for the parallel processing of 36 beams from a single observation.  
Below is a description of each step, including estimated processing time for a typical 10\,h observation (in brackets). 

\begin{description}

    \item[\textbf{Model creation} (\textit{15--20\,h})] 
    The sky model is created using calibrated visibility data. 
    See description in Section~\ref{subsec:model_images}. 

    \item[\textbf{Model subtraction} (\textit{a few minutes})]
    The model visibilities are subtracted from the original calibrated visibilities. 

    \item[\textbf{Short image creation} (\textit{1\,h})]
    A series of model-subtracted snapshot images (15-min time-step in this work) are generated using the model-subtracted visibilities. 
    See description in Section~\ref{subsec:snapshot_images}. 

    \item[\textbf{Source detection} (\textit{a few minutes})] 
    Untargeted source finding is run on model images. 
    See description in Section~\ref{sec:variability_search}. 

    \item[\textbf{Light-curve creation} (\textit{$<$1\,min})]
    For each source detected in model images, a radio light-curve is formed by collecting flux measurements from all model-subtracted snapshot images. 
    See description in Section~\ref{subsec:lightcurve_analysis}. 

    \item[\textbf{Statistical map analysis} (\textit{10--30\,min})]
    The model-subtracted snapshot images are used to form several statistical maps for further variability analysis. 
    See description in Section~\ref{subsec:statistical_map}. 

    \item[\textbf{Transient detection} (\textit{$<$1\,min})]
    The generated light-curves and statistical maps are analysed to detect transient or variable behaviour. 
    See detection criteria in Section~\ref{sec:variability_search}. 

    \item[\textbf{Automatic Filtering} (\textit{$<$1\,min})]
    Candidates that meet metrics listed in Section~\ref{subsec:lightcurve_analysis} are automatically discarded. 
\end{description}

The final data products are candidates catalogues, and model image cutouts, animations made from model-subtracted snapshot images for each transient candidate. 
These final products are transferred to the VAST local machine for manual inspection and further investigation, described as follows: 

\begin{description}
    \item[\textbf{Manual inspection}]
    The candidates are manually inspected to confirm if they are real transients.

    \item[\textbf{Multi-wavelength cross-matching}]
    Sources that exhibit transient or variable behaviour are cross-matched with multi-wavelength catalogues to investigate their nature. 
    See details in Section~\ref{sec:results}. 

    \item[\textbf{Transient classification}]
    Sources that exhibit transient or variable behaviour are classified based on their multi-wavelength information and/or other follow-up observations.
    See details in Section~\ref{sec:results}
    
\end{description}

In summary, the overall processing time for a 10\,h observation (36 beams in parallel) is about 1--2 days to get final transient candidates. 
The processing time is mainly limited by the model creation process. 
We discuss possible ways to improve this in Section~\ref{subsec:future_plans}. 
A detailed description of our dedicated data processing parameters is given in Section~\ref{subsec:data_processing}.

\subsection{Processing parameters}
\label{subsec:data_processing}

Each observation was calibrated and processed following the standard procedure (see detailed description in each survey paper, e.g., \citealt{Norris2021,Koribalski2020,Dobie2019}). 
After calibration, flagging, and self-calibration, the processed visibilities for each observation (total of 36 visibility sets, one for each of 36 beams) were uploaded to CASDA. 
These calibrated visibilities were used for our dedicated processing described as below. 
We processed each beam independently. 

\subsubsection{Model images}
\label{subsec:model_images}

We reduced the data using the Common Astronomy Software Applications package (\textsc{casa}; \citealt{McMullin2007}). 
For individual beams we created independent model images using \textsc{casa} \texttt{tclean} task from the self-calibrated visibilities. 
We used multi-scale multi-frequency synthesis with two Taylor terms \citep{Rau2011} to allow modelling the spectral curvature during the deconvolution process to obtain a better model at the reference frequency. 
We performed a deep clean with 10,000 iterations using Briggs weighting (robustness of 0.5) to obtain a good balance between resolution and sensitivity \citep{Briggs1995}. 
We chose an image cell size of 2.5\,arcsec and a large image size of 10,000$\times$10,000 pixels to include as many of neighbouring (bright) sources into the model, reducing side-lobes effects.
The typical residual rms is about 50\,\ujy for observations at low-frequency band and 35\,\ujy\ for observations at mid-frequency band. 
Future improvements would include flexible pixel size settings (for observation at different frequencies) and better handling of nearby bright sources (e.g., `A' sources peeling). 
For each observation we obtained 36 separate model images. 

\subsubsection{Model-subtracted snapshot images}
\label{subsec:snapshot_images}

We converted each model image to model visibilities, and subtracted the model visibilities from the original self-calibrated visibilities. 
We imaged the resulting model-subtracted visibilities in 15-min time-steps using the same weighting parameters as above. 
We chose a square image size of 3,000 pixels ($\sim$2.1\degr), which is approximately 1.5 times the primary beam diameter at 943.5\,MHz. 
We did not apply any deconvolution during this step as the model has already been subtracted from the calibrated visibilities. 
We did not perform any primary beam effect correction either as we were only interested in the relative flux densities of source variations along with time. 
For a typical 10\,h observation, we generated 40 model-subtracted snapshot images for each beam. 
The median rms noise for each model-subtracted snapshot image is about 200\,\ujy.

\section{Variability search}
\label{sec:variability_search}

We conducted a variability search using the model-subtracted 15-min snapshot images. 
Our search contains two different analysis methods. 
First, we did light-curve analysis focusing on characterising light-curves of all sources detected in the deep model images. 
Second, we did image analysis focusing on producing statistical maps (e.g. chi-square map and peak map) using all model-subtracted snapshot images and selecting any candidates that stood out in the map. 
The image analysis used a slightly stricter threshold to select candidates compared to the light-curve analysis as we do not use prior information of source positions in the searching process, but the advantage is being able to find any extreme transients (e.g. a single flare) that were not detected in the model images. 
We describe each of these approaches below.

For both methods a source catalogue from the model image was necessary as a reference to rule out false candidates. 
We used \textsc{aegean} \citep{Hancock2012,Hancock2018} to perform source finding at a $6\sigma$ threshold in model images and the built-in package \textsc{bane} to estimate the background and rms noise levels. 
We detected approximately 1,000 sources in each model image (per beam). 
Note that the model image size is about $2.1\times2.1$\,deg$^2$ and therefore there is an overlap for model images of neighbouring beams (beam spacing is 0.9\degr), which means most of sources are detected at least twice in neighbouring model images -- this is useful for checking the reliability of any measured variations. 

\subsection{Light-curve analysis}
\label{subsec:lightcurve_analysis}

For each source detected in the model image of a given beam, we converted the global coordinate to the pixel position in each image, and extracted its light-curve as follows:
\begin{enumerate}
    \item measured the flux density in the deep restored model image $S_\mathrm{deep}$ at the given pixel position; 
    \item measured the flux density on the $i$-th snapshot images $S_{i,\mathrm{snap}}$ at given pixel position;
    \item measured the residual flux density $S_\mathrm{residual}$ on the sky model residual image; and
    \item the $i$-th data point of the light curve is then given by 
    \begin{equation}
        S_i = S_\mathrm{deep} + S_{i,\mathrm{snap}} - S_\mathrm{residual}
    \end{equation}
\end{enumerate}

For each light-curve, we calculated the weighted reduced $\chi^2$ (defined as $\eta$) to measure the significance of random variability using 
\begin{equation}
    \eta = \frac{1}{N-1} \displaystyle\sum_{i=1}^{n} \frac{\left(S_i - \bar{S}\right)^2}{\sigma_i^2}
    \label{eq:chisquare}
\end{equation}
where $N$ is the total number of measurements in the light-curve (i.e., the number of model-subtracted snapshot images), $\sigma_i$ is the local rms measured at $i$-th snapshot image using \textsc{bane}, and $\bar{S}$ is the weighted mean flux density defined as 
\begin{equation}
    \bar{S} = \frac{\sum_{i=1}^{n} \left(S_i / {\sigma_i^2}\right)}{\sum_{i=1}^{n} \left({1} / {\sigma_i^2}\right)}
\end{equation}
As discussed in \citet{Rowlinson2019}, the histogram of $\eta$ for all sources follows an approximately Gaussian distribution in logarithmic space. 
Since all stable sources will follow this distribution, any sources with excessively large $\eta$ are considered to be variable sources. 
In this paper we selected candidates with $\eta>2.0\sigma_\eta$ where $\sigma_\eta$ is the standard deviation measured by fitting a Gaussian function to the $\eta$ distribution. 

We also calculated the weighted modulation index to measure the magnitude of variability using 
\begin{equation}
    m = \frac{\sigma_\mathrm{s}}{\bar{S}}
    \label{eq:modulation_index}
\end{equation}
where $\sigma_\mathrm{s}$ is the standard deviation of flux densities of the light-curve. 
We considered any candidates with $m \leq 5\%$ as non-variable. 
We ruled out extended sources (using the integrated-to-peak criterion below) as the varying ($u,v$) coverage during the observation can result in the measured flux density varying in different parts of the extended source. 
We also ruled out candidates near the edge of the beam. 

\begin{figure}
    \includegraphics[width=\columnwidth]{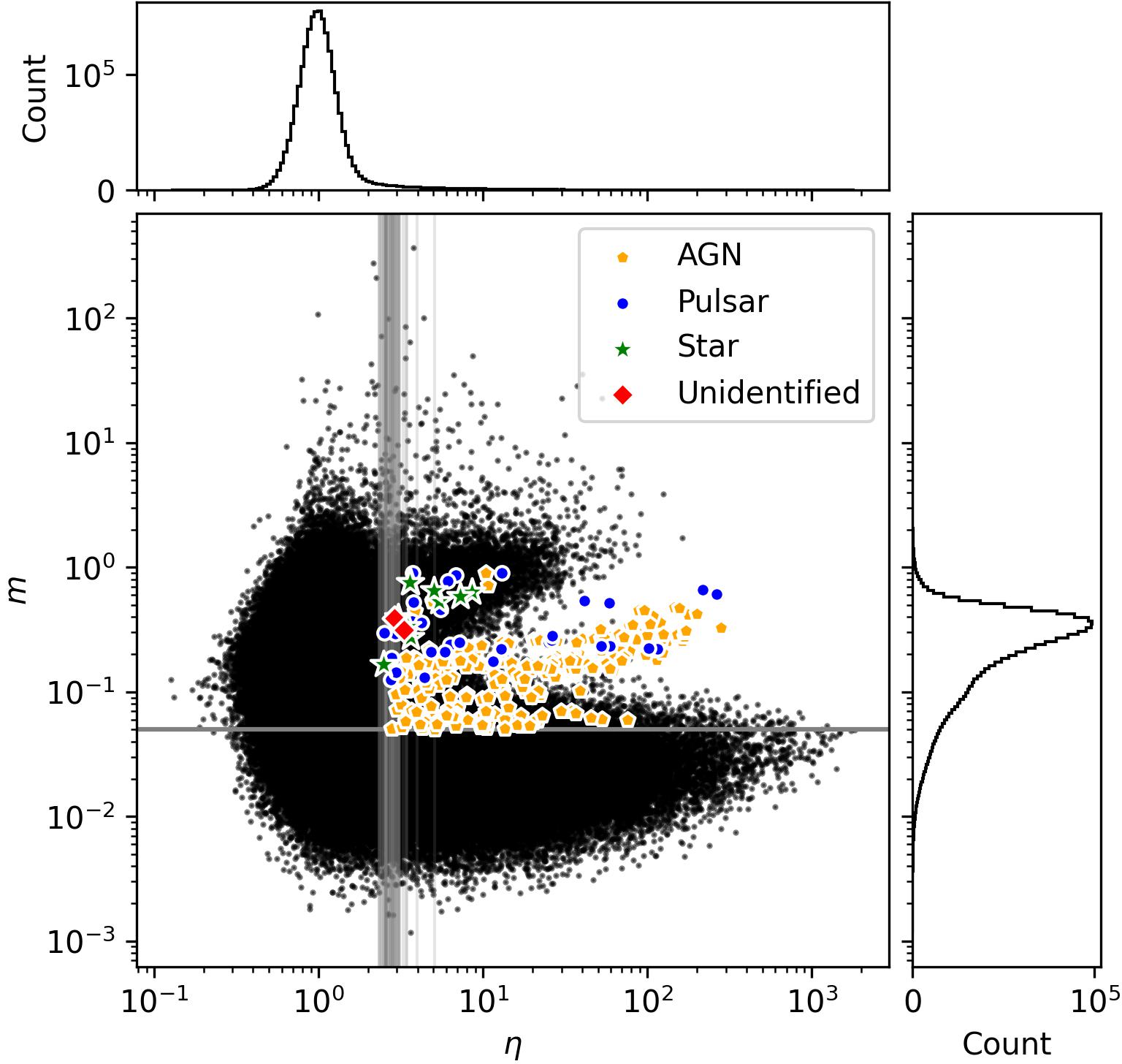}
    \caption{Distribution of the two key variability metrics, $\eta$ and $m$, for all sources in the light-curve analysis. The grey lines represent the selection thresholds (i.e., $m>0.05$ and $\eta>2.0\sigma_\eta$). The thresholds for $\eta$ is around 2.5--3.0 for the majority of fields. The colourful markers represent classified variable and transient sources: AGNs (orange pentagons), pulsars (purple circles), stars (green stars), and unidentified objects (red diamonds). See details in Section~\ref{sec:results}. Each unique source corresponds to multiple light-curves
    detections from neighbouring beams and/or overlapped fields. Another population of detections above the threshold (at $m\sim$1) are mostly false candidates near bright sources. }
    \label{fig:metrics} 
\end{figure}

Figure~\ref{fig:metrics} shows the distribution of sources in the variability metrics $\eta$ and $m$, defined in equation~(\ref{eq:chisquare}) and (\ref{eq:modulation_index}). 
The $2\sigma_\eta$ threshold corresponds to $\eta\sim2.5$--3.0 for the majority of fields. 
We noticed that objects in SB11832 has excessively large variability metrics compared to other fields (e.g., with a median $\eta\sim$2.5 and $2\sigma_\eta \sim$11). 
After investigation this field contains a Jy-level source PKS~1613$-$586, and its side-lobes largely affect many beams in this field. 
We discarded SB11832 in the following analysis. 

In summary, we selected candidates based on the following metrics:
\begin{enumerate}
    \item $\eta>2.0\sigma_\eta$ where $\sigma_\eta$ is measured from the $\eta$ distribution for all sources in each beam; 
    \item $m>0.05$; 
    \item ratio of integrated to peak flux density $<1.5$; 
    \item located within about one full-width half maximum (FWHM) from the beam centre, i.e., $<0.8$\,deg for observations at 887.5\,MHz and 943.5\,MHz and $<0.6$\,deg for observations at 1367.5\,MHz.   
\end{enumerate}
All candidates that passed the above metrics were subject to further manual inspection to rule out artefacts near bright sources and/or caused by rotated ($u.v$) sampling. 
We identified a total of 35 unique variable sources (as shown in the top-right quadrant of Figure~\ref{fig:metrics}). 
We compared the lightcurve of each source with that in neighbouring beams to confirm that the variability is genuine.  

\subsection{Statistical map analysis}
\label{subsec:statistical_map}

\begin{figure}
    \includegraphics[width=\columnwidth]{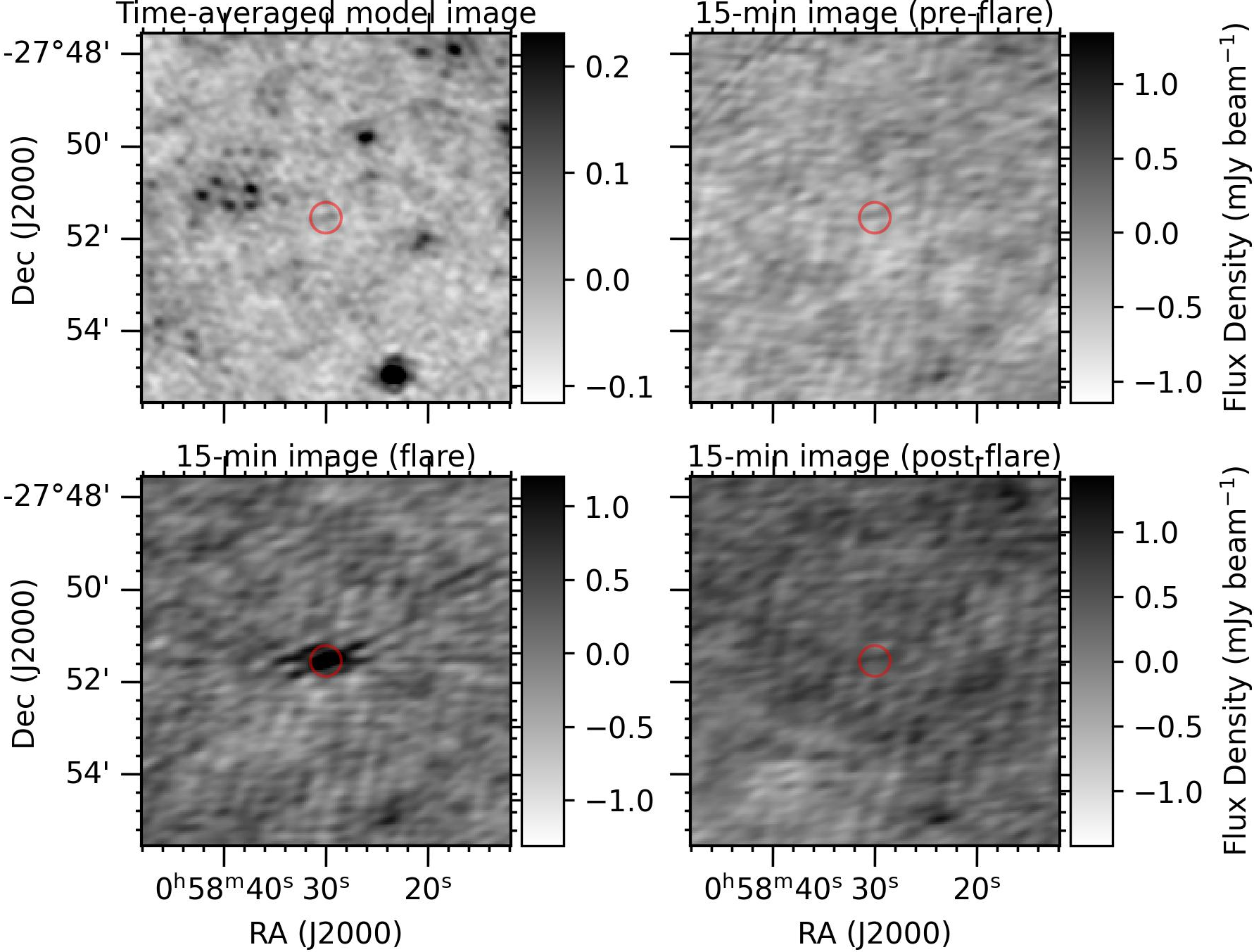}
    \caption{Time-averaged model image (top left) and 15-min model-subtracted snapshot images before flare (top-left panel), during flare (bottom left), and post flare (bottom right) of the single flare event VAST~J005830.0$-$275132 (CD-28~302) in SB12704. This object was not detected in the model image, but has a $12\sigma$ detection in one 15-min snapshot image. The radial spikes (bottom left) are from the not-deconvolved point spread function. }
    \label{fig:snapshot}
\end{figure}

The light-curve approach is sensitive to all variable sources detected in the model images. 
However, it cannot identify transient events that were not detected in an hours-long model image (e.g., single flare events; see an example in Figure~\ref{fig:snapshot}).

Hence we used a different approach generating different statistical maps (e.g., chi-square map, peak map, and Gaussian map) made from model-subtracted snapshot images, in order to identify transient phenomena independent of whether they were detected in model images. 
We first generated a cube for each beam using a series of model-subtracted snapshot images. 
For a 10-h observation, the resulting cube size in pixel is $[40,3000,3000]$. 
The first axis is time (a total of 40 15-min images for a 10-h observation), and the other two axes are the spatial dimensions (a size of 3,000 pixels on each dimension). 
For each spatial pixel ($x_p,y_p$) we extracted its flux density in $i$-th snapshot image to form a light-curve series $S(x_p, y_p)$.  
We generated 2D statistical maps using measured $S(x_p, y_p)$ as discussed below. 

\begin{description}
    \item[\textbf{Chi-square map}]
    We calculated the weighted reduced chi-square at each spatial pixel $\eta{(x_p,y_p)}$ based on extracted $S(x_p, y_p)$ using equation~(\ref{eq:chisquare}). 
    The chi-square map is therefore made up of $3000\times3000$ measured $\eta{(x_p,y_p)}$ values.
    For pixels with no, or weak, variability along the time axis, the resulting $\eta{(x_p,y_p)}$ should be around 1.  
    As defined in equation~(\ref{eq:chisquare}), the chi-square map is sensitive to any random variability, but less sensitive to a single flare event.
    
    \item[\textbf{Peak map}] 
    The peak map is the peak value measured in the time series for each ($x_p,y_p$). 
    Instead of using the measured flux density $S(x_p, y_p)$ directly, we calculated the signal-to-noise ratio $\mathrm{SNR}=S_{i,\mathrm{snap}}/\sigma_i$ at each ($x_p, y_p$). We selected the peak SNR($x_p, y_p$) along time dimension. 
    The advantage of using SNR is to account for local rms variation across the spatial image, e.g., the local rms around bright sources is relatively higher and therefore it has larger chance to produce a high noise spike. 
    The peak map is sensitive to variables with a high maximum flux density (e.g., single flare events). 
    
    \item[\textbf{Gaussian map}] 
    We created a 1D Gaussian filter kernel with standard deviation of 4 time-steps. 
    We convolved the light-curve $S(x_p, y_p)$ with the Gaussian kernel and stored the maximum value of the convolved light-curve at each spatial pixel. 
    The Gaussian map is made up of maximum values measured along time dimension at all spatial pixels, with a size of $3000\times3000$. 
    As our filter kernel has a standard deviation of 4 time-steps, the Gaussian map is sensitive to $\sim$1\,h long Gaussian-shape pulsed objects. 
    For a longer-timescale variable it should be detected in the chi-square map, and a short-timescale transient should be identified in the peak map. 
    Our Gaussian map can therefore fill the gap between the other two maps. 
\end{description}

\begin{figure}
    \includegraphics[width=\columnwidth]{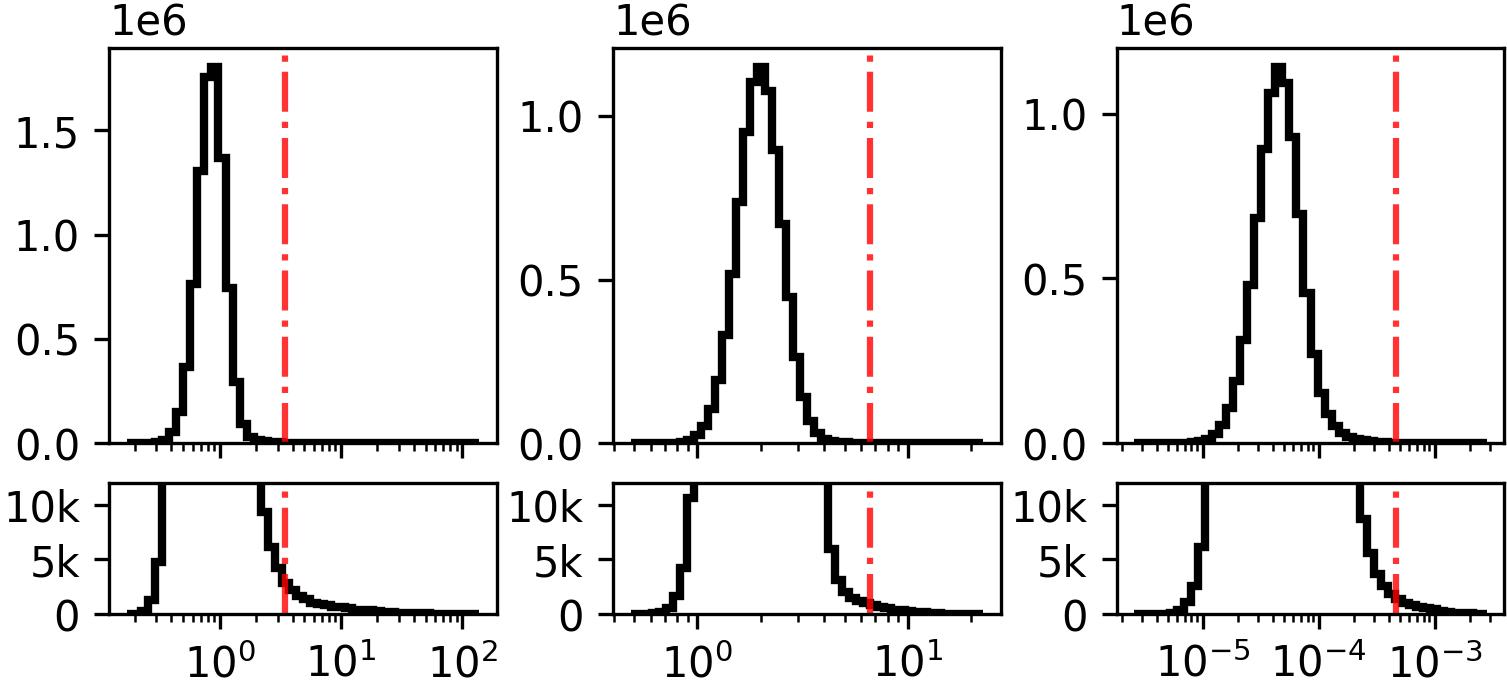}
    \caption{Distribution of the chi-square map (left), peak map (middle), and Gaussian map (right) in logarithmic space. The red dash-dot line represents the selection threshold. The bottom panel is a zoomed-in version showing the tail of the distribution. }
    \label{fig:statistical_maps}
\end{figure}

After we produced a chi-square map, peak map, and Gaussian map for each beam, we selected all local maxima above a certain threshold.
We used a relatively conservative threshold equal to 5 times the rms level measured in distributions of individual maps in logarithmic space to suppress the effects of extreme outliers (e.g., artefacts around bright sources). 
Figure~\ref{fig:statistical_maps} shows the distribution and the selection threshold of three maps from an example field. 
We noticed that these maps consist of correlated spatial pixels, and this may affect our statistics of threshold selection. 
We checked the distributions of values from every 4 (and 8) pixels in maps of a randomly selected field, and the results do not change.
We set a relatively strict limit for the separation of two local maxima $>30$ pixels (75\,arcsec) to reduce false detections due to nearby bright sources. 
We used the coordinates for each local maximum to cross-match with the model image catalogue. 
If this local maximum was isolated and not near to any sources in the model image catalogue (i.e., separation $>$30\,arcsec), the coordinates were directly stored in the candidate list. 
Otherwise we calculated the modulation index $m$ following equation~(\ref{eq:modulation_index}), and only selected candidates that meet the metrics described in Section~\ref{subsec:lightcurve_analysis} (i.e., objects with large $m>0.05$, compact, and close to beam centre). 
We combined the candidates selected from three maps using a 10\,arcsec cross-matching radius, and generated a final candidate list for each beam. 
All objects in the final candidate list were manually inspected to rule out artefacts. 
We identified a total of 36 highly variable and transient sources from this analysis, and three of which were not detected in the light-curve analysis. 
These objects (VAST~J005830.0$-$275132, VAST~J044649.5$-$603408, and VAST~J104918.8$-$250924) are single flare events that were only detected in one or two 15-min model-subtracted snapshot images each (see an example in Figure~\ref{fig:snapshot}). 

\section{Results} 
\label{sec:results}

We found 35 variable sources using the light-curve analysis approach and 36 variables and transient sources using the statistical map analysis approach, giving a total of 38 unique sources as listed in Table~\ref{tab:candidates}. 
Their light-curves are shown in Figure~\ref{fig:lightcurves}. 
Note that some observations overlapped and therefore one unique object was detected in multiple observations and has more than one light-curve. 
The flux density $S_\textrm{deep}$ listed in Table~\ref{tab:candidates} is measured in the deep model image without primary beam correction. 
In the following discussion we also used flux densities from catalogues of standard ASKAP outputs in CASDA, which are measured from primary beam correct, mosaiced image using \textsc{selavy} source finder \citep{Whiting2012}.
We made a note when the flux density is from the \textsc{selavy} catalogue. 

To do an initial classification we looked up each object in the SIMBAD Astronomical Database\footnote{\url{http://simbad.cds.unistra.fr/simbad/}} and the ATNF Pulsar Catalogue\footnote{\url{https://www.atnf.csiro.au/people/pulsar/psrcat/}} \citep{Manchester2005} using a cross-match radius of 2\,arcmin (considering possible proper motion). 
We searched archival multiwavelength data using a cross-match radius of 10\,arcsec from 
the Wide-field Infrared Survey Explorer (\textit{WISE}; \citealt{Wright2010}), 
the Two Micron All Sky Survey (2MASS; \citealt{Skrutskie2006}), 
\textit{Gaia} \citep{GaiaCollaboration2018}, 
the Dark Energy Survey (DES; \citealt{Abbott2018}), 
and the DESI Legacy Imaging Surveys \citep{Dey2019}.   
We applied positional corrections for query results using proper motion parameters from \textit{Gaia} DR3 (when available) or SIMBAD (normally for pulsar counterparts) to make sure the corrected position is within our positional uncertainty. 
We also generated image cutouts (2\,arcmin size) from multi-wavelength data (\textit{WISE} and 2MASS) and inspected radio overlays to confirm the counterparts are matched. 
We checked previous radio surveys including the GaLactic and Extragalactic All-sky Murchison Widefield Array (GLEAM; \citealt{Hurley-Walker2017}) at 200\,MHz, the Sydney University Molonglo Sky Survey (SUMSS; \citealt{Mauch2003}) at 843\,MHz, and the Rapid ASKAP Continuum Survey (RACS; \citealt{McConnell2020,Hale2021}) at 888\,MHz. 
Seven of our sources were found to be known pulsars and eight of them are stellar objects. 
22 are likely associated with AGNs or galaxies, and one are unidentified.

For each source we produced full Stokes dynamic spectra with the highest possible time and frequency resolution (10\,s and 1\,MHz) to investigate the variability behaviour in more detail. 
We generated dynamic spectra using model-subtracted visibilities (see Section~\ref{subsec:data_processing}). 
We first phased rotated the model-subtracted visibilities from the beam centre to the coordinates of the target object. 
We averaged the visibilities for each instrumental polarisation only using baselines longer than 200\,m (to exclude diffuse emission). 
We then combined the complex visibilities to generate dynamic spectra with full-Stokes parameters. 
Below we discussed these variables and transient sources with information from multi-wavelength search and dynamic spectra.

\afterpage{
\input{table_candidates.tex}
\input{figure_lightcurves.tex}
}

\subsection{Pulsars}

Seven of our highly variable sources are known pulsars, of which three are millisecond pulsars (MSP) in binary systems (see Table~\ref{tab:candidates}). 
Notes on some highlighted pulsars are given below.

\begin{figure}
    \includegraphics[width=\columnwidth]{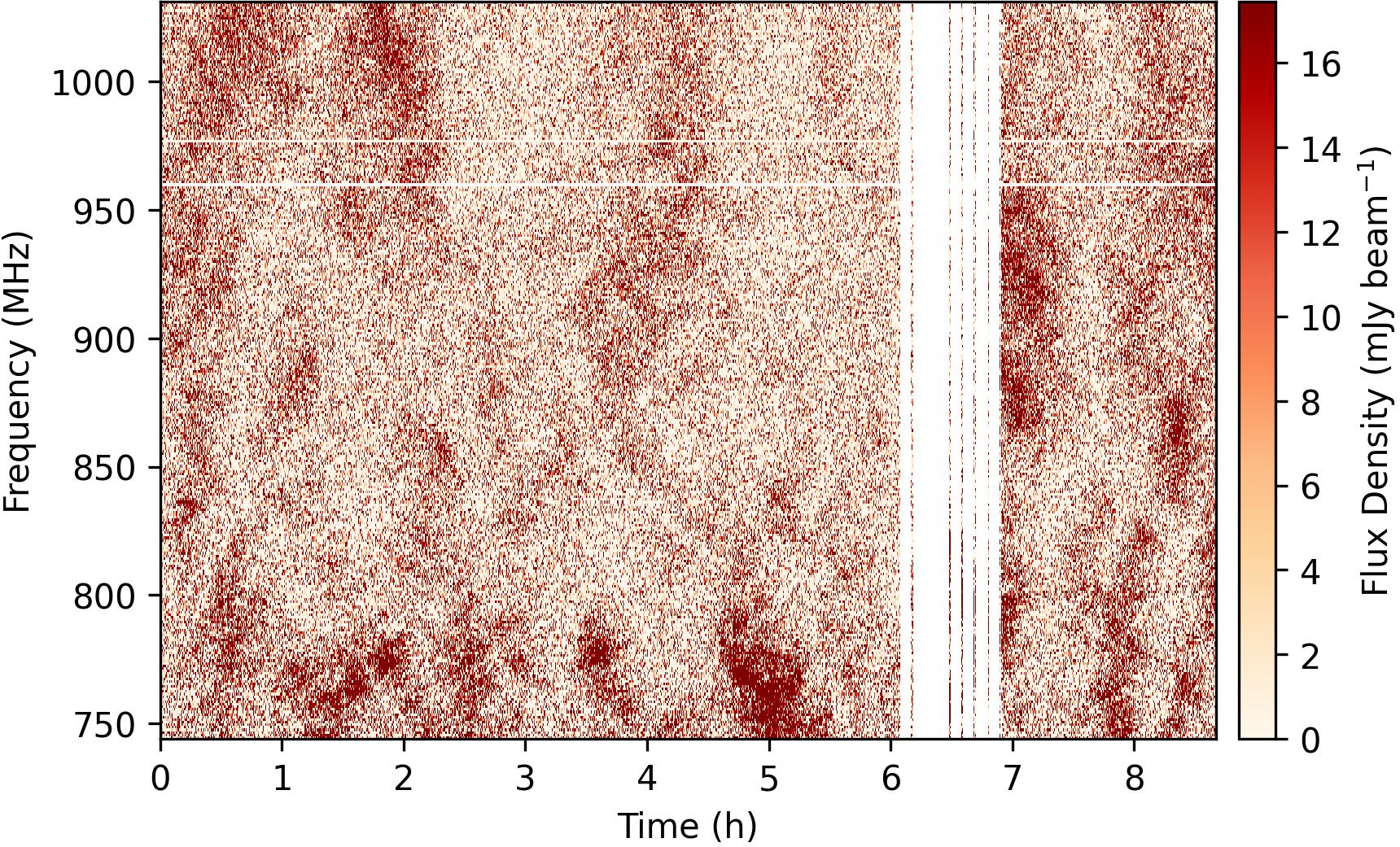}
    \caption{An example dynamic spectrum from one of identified pulsars (PSR~J0837$+$0610 in SB10123). The narrow-band scintles are clearly shown. The flagged data are represented as white colour. }
    \label{fig:scintles}
\end{figure}

\begin{description}

\item{\bf PSR~J0837$+$0610} and {\bf PSR~J0922$+$0638} are two bright and well-studied pulsars. 
Both of them have significant diffractive scintillation reported in literature \citep[e.g.][]{Cordes1986,Bhat1998,Thyagarajan2011,Wu2022}. 
PSR~J0837$+$0610 is also one of the first known nulling pulsars noticed by \citet{Backer1970}. 
We observed narrow-band scintles in the dynamic spectra (Figure~\ref{fig:scintles}), consistent with what 
\citet{Cordes1986} reported at a similar frequency (1000\,MHz). 

\item{\bf PSR~J1704$-$6016} has a currently known position of
RA = 16$^\mathrm{h}$59$^\mathrm{m}$47\fs9, 
DEC = $-$60\degr12\arcmin43\arcsec (B1950) from timing observations with the Parkes telescope \citep{Newton1981}.  
Although this pulsar is about 2.7\,arcmin away from VAST~J170416.8$-$601934, we suggest there is an association between them. 
VAST~J170416.8$-$601934 was detected as 24.4\,mJy in SUMSS (843\,MHz), 10.6\,mJy in RACS (888\,MHz), and 2.3--4.0\,mJy (from the \textsc{selavy} catalogue) in our data (SB12193, SB12209; 1367.5\,MHz). 
It also shows a faint counterpart in GLEAM (200\,MHz). 
According to this VAST~J170416.8$-$601934 is presumably a steep-spectrum source, with a crude estimation of spectral index $\alpha\sim-2$ ($S\propto\nu^\alpha$). 
VAST~J170416.8$-$601934 has no counterpart at other wavelengths including infrared, optical, or $\gamma$-ray. 
We observed narrow-band scintles in the dynamic spectra, suggesting its compact nature. 
All of above properties are consistent with a pulsar. 
PSR~J1704$-$6016 was reported as 23\,mJy at 400\,MHz by \citet{Taylor1993}. 
However, there is no source at its catalogued position from any of above radio continuum surveys. 
We therefore suggest there is a position error in the original pulsar catalogue, and PSR~J1704$-$6016 is the counterpart of VAST~J170416.8$-$601934.  

\item{\bf PSR~J2039$-$5617} is a known $\gamma$-ray source and was recently confirmed as a 2.6\,ms radio pulsar with orbital period of about 5.5\,h \citep{Clark2021,Corongiu2021}. 
This orbital modulation causes radio emission obscured for about half the orbit -- roughly consistent with our light-curve in Figure~\ref{fig:lightcurves} (covering a full orbit phase range with a rising trend at the end of the observation SB9351). 
PSR~J2039$-$5617 shows significant variation in pulsed amplitude and has been detected in the VAST Phase I Pilot Survey \citep{Murphy2021} as a highly variable source on longer timescales ($\sim$days). 



\item{\bf PSR~J2144$-$5237} is a 5\,ms pulsar in a binary system (orbital period of $\sim$10 days) identified by \citet{Bhattacharyya2016,Bhattacharyya2019} using the Giant Metrewave Radio Telescope (GMRT). 
They suggested a ‘redback’ system for this object based on companion mass, but found no sign of eclipses in their timing data. 
PSR~J2144$-$5237 was detected in two of our datasets (SB9434 and SB10168), showing clear frequency structure in the dynamic spectra. 
It has broad-band scintles (comparable with the observing bandwidth of 144\,MHz) at SB10168 (1367.5\,MHz) and relatively narrow-band scintles at SB9434 (943.5\,MHz). 
This object has also been detected in VAST Phase I Pilot Survey. 

\end{description}

In summary, we observed narrow-band scintles in the dynamic spectra from six of seven pulsars (see Figure~\ref{fig:scintles} as an example). 
The remaining object, PSR~J2039$-$5617, is an eclipsing MSP. 
Pulsars are generally most extreme variables in our sample, with higher $\eta$ and $m$ than most of stars and AGNs/galaxies (as shown in Figure~\ref{fig:metrics}). 

\subsection{Stars}
\label{subsec:stars_results}

\input{table_stars}


Eight of our identified transients are radio stars. 
Five of the stars are M dwarfs and one (CD-51~13128) is an M dwarf binary. 
One (\replace{$\lambda$ Columbae}) is classed as a B5V star. 
The other star (UPM~J1709$-$5957) do not have classifications. 
Six of the eight stars have previously been identified as X-ray stars, while the other two are coincident with X-ray sources. 
Here we provide a brief description of each source.

\begin{description}

\item{\bf CD-28~302} is a single-line spectroscopic binary \citep{2018MNRAS.475.1960F} where the main component is a dM4 star \citep[e.g.][]{2019AJ....158...56H}. It was identified as a non-single star in \textit{Gaia} DR3 with an orbital period of $\sim$147.0\,days \citep{GaiaMission,GaiaDR3}.
The source has not been definitely identified as an X-ray source; however, there are detections of an X-ray source within a few arcseconds of the source position by \textit{ROSAT} \citep{1999A&A...349..389V}, \textit{XMM-Newton} \citep[e.g.][]{2020A&A...641A.137T} and \textit{Chandra} \citep{2010ApJS..189...37E}.
CD-28~302 has not previously been identified as a radio star.
The \textit{Gaia} DR3 parallax is $84.98\pm0.46$\,mas, with a distance of $11.77^{+0.07}_{-0.06}$\,pc \citep{GaiaEDR3dist}. 

\item{\bf BPS~CS~29520-0077} is a single M1.5V star \citep[e.g.][]{2006AJ....132..866R,2014ApJ...788...81M}.
It was noted to have strong Calcium H and K emission lines by \citet{1996AJ....112.1188B} and is a known X-ray star \citep{2018A&A...614A.125F}.
This object was detected in the radio by \citet{Rigney2022} using ASKAP. They measured a peak Stokes I flux density of $0.25\pm0.01\,\mathrm{mJy\,beam^{-1}}$ over the full 13 hour integration, and identified the same radio peak we show here after dividing the 13 hour observation into 10 minute slices. \citet{Rigney2022} used the same procedure as described in this work to investigate the short-timescale light curve.
The \textit{Gaia} DR3 parallax is $40.73\pm0.01$\,mas, with a distance of $24.542^{+0.006}_{-0.008}$\,pc \citep{GaiaEDR3dist}.


\item{\bf \replace{$\lambda$ Columbae}} is a variable B5V star \citep[e.g.][]{1930MmMtS...2....1R,1975RMxAA...1..299J}. It was identified as a rotating ellipsoidal variable with a period of 1.3\,days by \citet{2006SASS...25...47W}, possibly indicating that \replace{$\lambda$ Columbae} is a tight binary. 
It is also a known X-ray star \citep{1999A&A...349..389V}.
\replace{$\lambda$ Columbae} has not previously been identified as a radio star.
The \textit{Gaia} DR3 parallax is $9.47\pm0.09$\,mas, with a distance of $105.4\pm0.9$\,pc \citep{GaiaEDR3dist}.

\item{\bf 2MASS~J10491880$-$2509235} is a single, young, lithium rich M4.9 star demonstrating strong H-alpha emission \citep{2015MNRAS.453.2220M}, and has not previously been reported as radio loud.
The star has also been associated with X-ray activity by \textit{ROSAT} \citep{1999A&A...349..389V}, \textit{XMM-Newton} \citep[e.g.][]{2020A&A...641A.137T} and \textit{Chandra} \citep{2010ApJS..189...37E} that is consistent with a young age \citep{2015MNRAS.453.2220M}.
The \textit{Gaia} DR3 parallax is $9.50\pm0.04$\,mas, with a distance of $104.6^{+0.4}_{-0.6}$\,pc \citep{GaiaEDR3dist}. 

\item{\bf UPM~J1709$-$5957} does not have a stellar classification and has not been previously reported as radio loud.
It is a known X-ray star, detected by \textit{ROSAT} \citep{2009ApJS..184..138H} and it was identified as a long-period variable by \citet{GaiaDR3}.
This source does not have a \textit{Gaia} parallax or distance measurement. The photometric distance was estimated to be $14.13\pm2.77$\,pc by \citet{2014AJ....148..119F}.

\item{\bf WT~713} is a photometrically classified M dwarf \citep{Sebastian2021} and has not been previously reported as radio loud. 
It was identified as an X-ray star using \textit{ROSAT} by \citet{2022A&A...664A.105F}.
The \textit{Gaia} DR3 parallax is $26.61\pm0.02$\,mas, with a distance of $37.49^{+0.04}_{-0.03}$\,pc \citep{GaiaEDR3dist}.

\item {\bf RX~J2138.5$-$5050} is an M7/8 ultra-cool dwarf candidate \citep{2022MNRAS.511.6179L} and an eruptive variable detected by \textit{TESS} \citep{Gunther2020}.
It was identified as an X-ray star using \textit{ROSAT} by \citet{2022A&A...664A.105F}.
The \textit{Gaia} DR3 parallax is $22.09\pm0.20$\,mas, with a distance of $44.96^{+0.3}_{-0.4}$\,pc \citep{GaiaEDR3dist}.

\item{\bf CD-51~13128} is an M dwarf binary with a 1.24\,day period \citep{1993MNRAS.260..132J}, with a third white dwarf component in a wide orbit around the M dwarf pair \citep[][]{1981AJ.....86..264W}.
It has been identified as an X-ray source using \textit{ROSAT} \citep{2009ApJS..184..138H,2022A&A...664A.105F}. 
The \textit{Gaia} DR3 parallax is $67.27\pm0.02$\,mas, with a distance of $14.858^{+0.004}_{-0.005}$\,pc \citep{GaiaEDR3dist}.

\end{description}

\begin{figure}
    \includegraphics[width=\columnwidth]{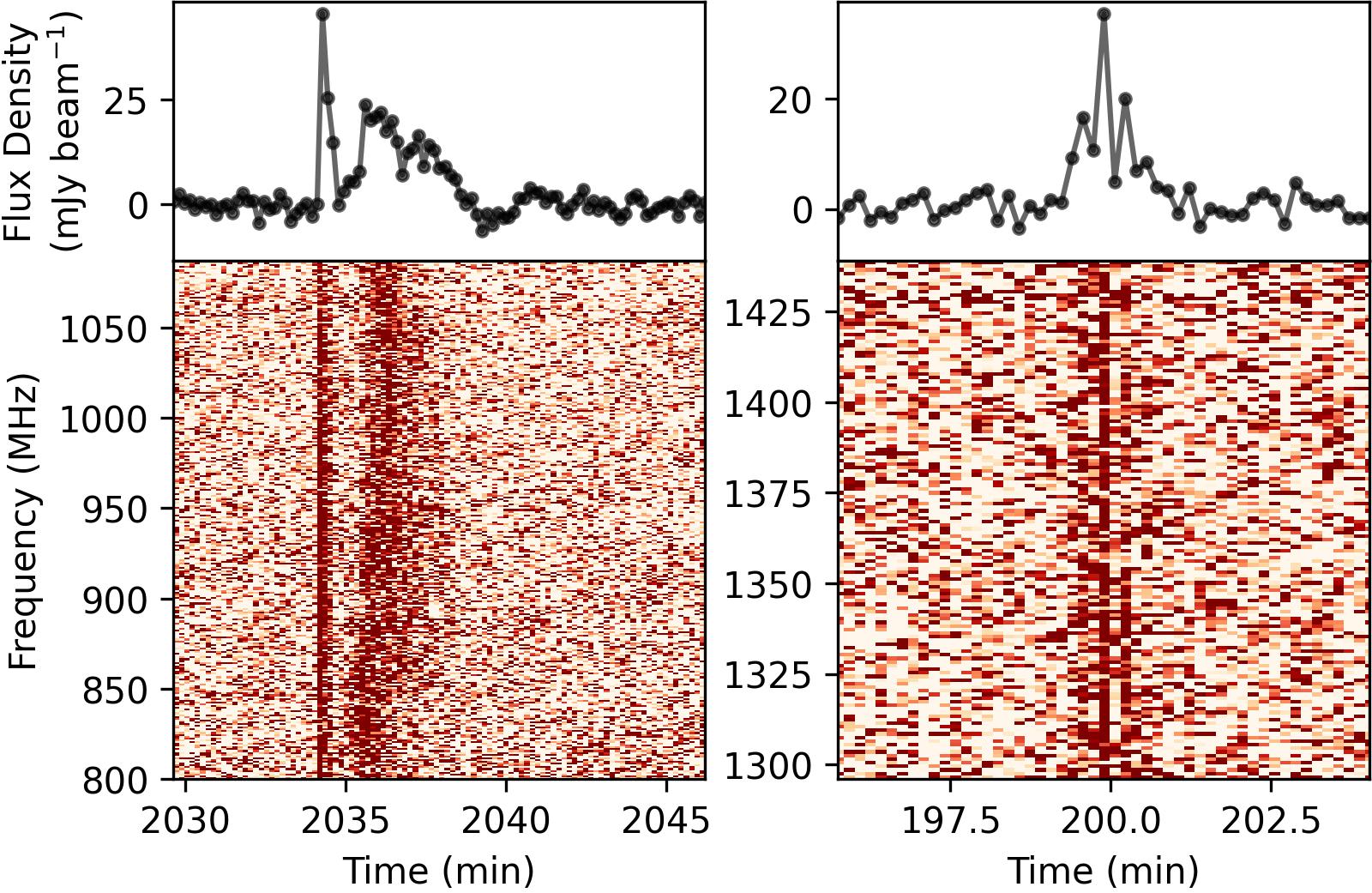}
    \caption{The total intensity dynamic spectra of two flaring stars CD-28~302 (left panel) and 2MASS~J10491880$-$2509235 (right panel). The temporal and spectral resolution are 10\,s and 1\,MHz, respectively. }
    \label{fig:flares}
\end{figure}

As shown in Figure~\ref{fig:lightcurves}, their light-curves have diverse morphologies and flare durations. 
After checked their dynamic spectra, two stars have even shorter flares than 15 minutes (see Figure~\ref{fig:flares}). 
One is CD-28~302 showing a $\sim$5\,min multi-component burst reaching a flux density of 45.6\,\mjy; 
another one is 2MASS~J10491880$-$2509235 showing a $100\%$ circularly polarised burst lasting about 1 minute and reaching a flux density of 35.5\,\mjy. 
\replace{
Table~\ref{tab:stars} lists the estimated variability timescales of all stars. 
They are approximate values identified by eye from dynamic spectra and 15-minute light-curves. 
The listed radio luminosities are calculated using the maximum flux density measured in the 15-minute light-curves or 10-second lightcurves if their variability timescales are shorter than 15 minutes. 
}
We present more discussions in Section~\ref{subsec:stars_variability}. 


\subsection{AGN/Galaxies}

We found 23 variable sources that were not identified as known pulsars or stars. 
For further investigation we checked the \textit{WISE} colours for those sources with a \textit{WISE} counterpart, 
the photometric redshift from the DESI Legacy Imaging Surveys DR8 catalog \citep{Duncan2022}, 
and \replace{the Million Quasars (Milliquas) catalogue \citep[v7.10;][]{Flesch2015,Flesch2021arXiv210512985F}}. 
Their multi-wavelength properties are summarised in Table~\ref{tab:candidates}. 

We found the majority of these objects (21) are associated with catalogued AGN or galaxies. 
The remaining two objects lack any obvious optical or infrared counterparts and are yet to be identified. 
One of these VAST~J045922.0$-$563419 was detected in SUMSS ($7.7\pm0.8$\,mJy) and hence is also likely associated with an AGN or galaxy. 
We present more discussions in Section~\ref{subsec:sky_distribution}. 

\subsection{An unclassified transient}
The remaining object, VAST~J205139.0$-$573906, has no multi-wavelength counterparts and was not detected in any archival radio surveys including GLEAM, SUMSS, and RACS. 
It was only detected in a single ASKAP observation SB9972, with a flux density of $0.58\pm0.03$\,\mjy\ (from the \textsc{selavy} catalogue) at 1367.5~MHz. 
This area has been covered by three other 10-hour ASKAP observations, and the non-detection implies a $3\sigma$ limit of 0.13--0.15\,\mjy\ at 943.5\,MHz (SB9325 and SB9351) and 0.17\,\mjy at 1367.5\,MHz (near the edge of SB9992). 
For the detected epoch SB9972, the intra-observation analysis shows a high variability ($m\sim$0.32) on timescales of miuntes to hours (see Figure~\ref{fig:lightcurves}) -- faster and more extreme than most variables associated with AGN/galaxies. 
VAST~J205139.0$-$573906 has strong linear polarisation of 38\%, with a rotation measure (RM) of about 8.4\,rad\,m$^{-2}$. 
It also has a $4\sigma$ peak ($0.13\pm0.03$\,\mjy) in the Stokes V image, suggesting a potential high fractional circular polarisation ($\sim$22\%) if the detection is real. 
Further observations are ongoing to determine the nature.

\section{Discussion}
\label{sec:discussion}

\subsection{The origin of variability in our sample}

\subsubsection{Pulsars}
\label{subsec:pulsars_variability}

For six of seven pulsars we observed narrow-band scintles in their dynamic spectra (as shown in Figure~\ref{fig:scintles}), with the only exception being PSR~J2039$-$5617 whose variability is presumably from eclipses. 
This shows that diffractive scintillation is the dominant reason of pulsar variability on this timescale at this observing frequency. 
This is not unexpected as refractive interstellar scintillation usually appears on much longer timescales (e.g., $\gtrsim$days).  
The presence of high quality scintles in the dynamic spectra also highlights the possibility of estimating screen distances (for pulsars with known proper motion) with ASKAP. 
More discoveries would improve our understanding of local scattering environment and the origins of scintillation. 

\replace{
From the ATNF Pulsar Catalogue, we found 14 pulsars that were detected in the 10-h deep ASKAP image but not selected by our pipeline, suggesting about one third of pulsars vary on 15-minute timescales based on our selection criteria. 
These ``non-varying'' pulsars generally have a much larger dispersion measure and fainter than our identified pulsars, making it harder to detect any variations. 
Future dedicated investigation is required to quantify pulsar behaviours on short timescales. 
}

\subsubsection{Stars}
\label{subsec:stars_variability}

The majority of our identified stars are M-dwarf stars, which are known to produce radio flares on timescales ranging from seconds to hours due to strong magnetic fields -- similar to what we detected \citep[e.g.,][]{Dulk1985,Villadsen2019}. 
\replace{
The radio luminosity of our identified stars ranges from $10^{14}$ to $10^{17}$\,erg\,s$^{-1}$\,Hz$^{-1}$ (see Table~\ref{tab:stars}), with the majority being around $10^{15}$\,erg\,s$^{-1}$\,Hz$^{-1}$. 
These values are consistent with other studies at similar observing frequencies \citep[e.g.][Pritchard et al. 2023, \textit{submitted}]{Villadsen2019,Pritchard2021}. 
}

In the left panel of Figure \ref{fig:flares} we show a Stokes I dynamic spectrum of a multi-component burst from CD-28~302 reaching a flux density of 45.6\,\mjy. 
The first component is elliptically polarised with $90\%$ circular polarisation and $10\%$ linear polarisation and lasts 30 seconds, while the second component is $100\%$ circularly polarised and lasts for about 4 minutes with a positive frequency drift rate of 3.3\,MHz\,s$^{-1}$. 
The high fractional circular polarisation and presence of elliptical polarisation suggests the emission is produced by the electron cyclotron maser instability (ECMI). 
The polarisation properties and burst morphology are similar to the Type IV burst detected from Proxima Centauri by \cite{Zic2020}, though with a significantly shorter delay between burst components.

We also detected a single, 100\% circularly polarised burst lasting $\sim$1 hour from \replace{$\lambda$ Columbae}, a B5V star. 
B-type stars are not typical radio emitters, lacking a magnetic dynamo mechanism and thus a means to produce the strong magnetic fields associated with highly circularly polarised radio emission. 
One possible explanation is that the emission originates from an undiscovered binary companion that is a more typical radio emitter. 
As mentioned in Section~\ref{subsec:stars_results}, \replace{$\lambda$ Columbae} is likely a tight binary, though \citet{Jerzykiewicz1993} found that the mass of any tidally influential companion must be less than $1 M_\odot$ and rules out tidal effects as a cause of variability. 
\replace{
We note that the binary system with a companion like this would have an extreme mass ratio $q\lesssim0.2$ -- uncommon in existing observations \citep{Moe2017ApJS..230...15M}. 
However, similar systems have been detected previously and provided valuable information for understanding stellar formations and evolution \citep[e.g.][]{Moe2015ApJ...801..113M}. 
The low detection number of extreme mass-ratio systems may also be a result of technique limitations. 
We therefore cannot rule out the presence of a sub-stellar companion in this system, and it is possible that a more typical radio emitter such as a K- or M-type dwarf may be the source of radio emission. 
We note that the estimated radio luminosity of $\lambda$ Columbae is consistent with other M-dwarf stars in our samples.
}
Another possibility is that \replace{$\lambda$ Columbae} is a magnetic chemically peculiar star that has not yet been identified. These stars are known to produce rotationally modulated pulses of circularly polarised radio emission \citep[e.g.][]{Trigilio2000, Leto2020, Das2022a} over timescales that align with the observed $\sim$1-h burst.

\subsubsection{AGN/galaxies}
\label{subsec:agn_variability}

For the 22 objects associated with AGN or galaxies, five of them are extreme IHVs discovered previously by \citet{Wang2021b}. 
We excluded these five known objects 
from this discussion. 

We first consider an intrinsic origin for the variability of the remaining 17 sources. 
If we consider the incoherent emission process, we can estimate the brightness temperature $T_b$ following the Rayleigh-Jeans law.
We cannot determine a reliable characteristic timescale since most of objects varies slowly in a monotonically increasing or decreasing way with a relatively low modulation index $\lesssim$10\%. 
Instead we consider the maximum flux density change $\Delta S$ in the corresponding time interval $\Delta t$. 
We obtain an expression for brightness temperature as follows: 
\begin{equation}
    T_b = \frac{d^2 \Delta S}{2k\nu^2 \Delta t^2}
\end{equation}
where $d$ is the distance estimated using photometric redshift (as listed in Table~\ref{tab:candidates}) in cosmology of $H_0=70$\,km\,s$^{-1}$Mpc$^{-1}$, 
$k$ is the Boltzmann constant and $\nu$ is the observing frequency. 
We found the estimated brightness temperature is about $10^{16}$--$10^{19}$\,K, greatly exceeding the inverse Compton limit $T_b\sim10^{12}$\,K for inconherent synchrotron emission \citep{Kellermann1969}. 
Even for a relativistic source (which is true for some AGNs), it would require a Doppler boosting factor $\sim$100 to explain the observed luminosity changes (Blazar normally has a factor of $\sim$10; \citealt{Hovatta2009}). 
We conclude that the observed variations are not likely to be intrinsic. 
This is not surprising given our observing frequency ($\sim$GHz) and timescale (15 minutes). 

If we consider the variability to have an extrinsic origin (i.e. propagation effects), we can calculate the expected scintillation transition frequency using NE2001 \citep{Cordes2002}. 
We then estimated the variability level from refractive scintillation based on \citet{Walker1998}. 
We find the expected point-source modulation index is $\sim$25\%--40\% and the scintillation timescale is $\sim$5--20 days for the lines-of-sight of these objects. 
This is much slower than what we observed, suggesting that normal scintillation caused by Kolmogorov turbulence in the diffuse, ionised ISM is unlikely to explain these variables. 

Considering the variability timescales we observed ($\sim$hours), the most likely explanation is enhanced scintillation caused by nearby plasma screens -- similar to (although slower than) the five known IHVs reported by \citet{Wang2021b}. 
The scintillation timescale for weak or refractive scattering increases with distance between the scattering screen and the observer \citep{Narayan1992}, and a relatively local scintillating screen can explain the variability timescales of these objects. 
For the origin of local scintillation screen, \citet{Walker2017} proposed that extreme scintillation may be associated with hot stars in the solar neighbourhood, and the scattering plasma is from thin skins on tiny molecular gas clouds surrounding the star. 
In our sample there are two variables offset by 20--30\,arcsec from two nearby stars (VAST~J042932.8$-$593118 and TYC~8515-1281-1; VAST~J053003.0$-$592700 and TYC~8530-1065-1). 
Both stars are $\sim$11\,mag at B band and $\sim$300\,pc away from us. 
We find a surface density of $\sim$0.001 per arcmin$^2$ for bright stars at this level, and would expect to find 0.001 stars within 1\,arcmin$^2$ area around a random object. 
However, we found 2 bright stars near these 22 objects with a separation of $<30$\,arcsec, which is two orders of magnitude higher than the expectated value. 
More observations (especially at different observing frequencies and on different date of year) are needed to constrain the screen kinematics and distance, and to confirm or rule out this association. 


\subsection{Sky distribution of AGN/galaxies}
\label{subsec:sky_distribution}

\begin{figure}
    \includegraphics[width=\columnwidth]{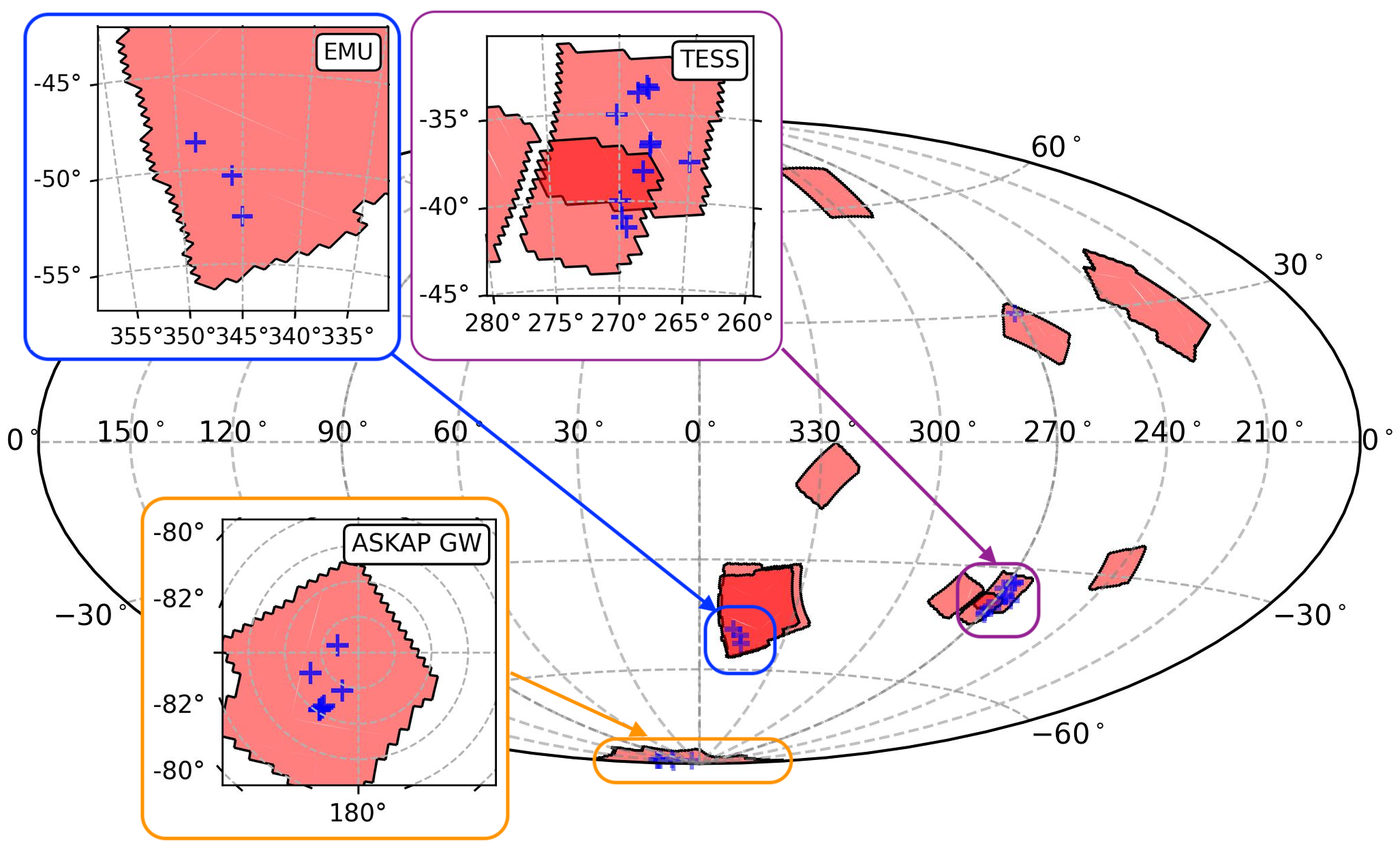}
    \caption{Sky distribution of scintillating AGNs we identified (shown as blue `+' marker). The sky map is plotted with Galactic coordinates, overlaid with red shadow representing the sky region covered by this survey. The zoommed-in insets show three regions clustered by scintillating AGNs. }
    \label{fig:scintillators}
\end{figure}

In the S190814bv field, \citet{Wang2021b} discovered five extreme scintillators in a linear arrangement in the sky, revealing a common long ($>$1.7\,degrees), thin (1--4\,arcmin), and straight plasma screen near to Earth (a few pc). 
This is the first time to known the geometry of a discrete, local cloud that causes extreme scintillation, raising a question of how common the similar plasma screen is in the solar neighbourhood. 

We searched a larger sky area in this work, and discovered 17 new scintillating objects. 
We found no obvious alignment for these new scintillators (see Figure~\ref{fig:scintillators}). 
\citet{Wang2021b} calculated the rate of recognising a similar filament is at least 16 per cent when discovered one scintillator. 
We therefore expect to recongnise at least 2 or 3 similar filamentary screens if they are responsible for the variations. 
Since we do not find any alignment, it is likely that these plasma screens are much shorter than the one found previously, or they are not in a straight filamentary shape. 
We detected two new variables
(VAST~J004711.9$-$262350 and VAST~J004929.7$-$241846)
in the original S190814bv field and noticed they are not located near or within the line. 
We analysed annual modulation of scintillation rates of the two new variables using the same method as \citet{Wang2021b}. 
The screen kinematics of the two variables are not well constrained as their variations are less extreme and slower. 
As a result, despite different best-fit values, given the large uncertainties the estimated kinematics are roughly consistent with each other and even consistent with that of the filamentary screen found previously.
It is possible that their variations are caused by the same scattering screen and the original filament is only part of a big structure, or their variations are from different origins. 

One interesting thing we noticed is the clustered sky distribution of 22 variables, as shown in Figure~\ref{fig:scintillators}. 
The total survey sky coverage is 1,476\,deg$^2$, but 21 of 22 objects are discovered in three relatively small areas: 
seven are in the S190814bv field, within a $\sim$10\,deg$^2$ area (near the south Galactic pole); 
11 are in the SB25077 and adjacent SB10636 fields, within a $\sim$50\,deg$^2$ area (galactic latitude $b$ from $-30$\degr to $-40$\degr); 
three are in the SB9434 and adjacent SB10635 fields, within a $\sim$15\,deg$^2$ area ($b$ from $-48$\degr to $-53$\degr). 
The only remaining object is VAST~J104933.6$-$245653 in the SB10612 field (around $b=30$\degr). 
This clustering is unusual, and at least implies that these regions have a greater abundance of turbulent gas clouds than other regions. 
Although we found no obvious association between these regions and H$\alpha$ intensity. 

\citet{Redfield2008} identified 15 local ISM clouds and proposed interactions between their boundaries as the origin of enhanced radio-wave scintillation. 
We found the locations of our variables are near to some of cloud-cloud interaction boundaries. 
For example, the 11 variables identified in TESS field (Figure~\ref{fig:scintillators}) are extremely close to the boundaries of the `G', `Vel', `Cet' clouds, and less favourably, the `Blue', and `Dor' clouds. 
Further observations are needed to establish the annual modulation of the scintillation rate
to determine the velocity of the screens along with the characteristic scale. 
This can then be used to compare with proposed clouds kinematics. 

Three clustered regions are about 75\,deg$^2$ out of 1,476\,deg$^2$ total sky coverage, we therefore expect that 5 per cent of sky are active regions with a number of extreme scintillators. 
Note not all compact objects within the active regions are scintillating, and the fraction of actual plasma screens should be much less than 5 per cent. 
Using our searching technique, we expect to detect more extreme scintillators in future all-sky surveys (e.g., EMU and WALLABY). 
This would give us a systematic view of all-sky distribution of local plasma screens, allowing further inspection of association with cloud-cloud interaction boundaries and other scenarios.

\subsection{Variability rate analysis}

We found 38 unique variable and transient sources in a sky area of 1,476\,deg$^2$ with total of 505\,h observations. 
We analysed 52 8--10\,h ASKAP fields and performed a variability search using 15-min model-subtracted snapshot images. 
Our variability search selected all sources detected in $\sim$10\,h model images and was therefore sensitive to any compact object with an averaged flux density of $\gtrsim$0.18\,mJy ($5\sigma$ rms threshold). 
A two-epoch equivalent sky coverage (see Equation~1 in \citealt{Bannister2011}) of our survey is about 88,140\,deg$^2$, which implies a variability rate of $4.3\times10^{-4}$ sources per two-epoch deg$^2$. 
If we exclude sources known to be Galactic (the stars and pulsars) this gives an surface density of $2.5\times10^{-4}$ sources per two-epoch deg$^2$. 
This value are consistent with predictions from \citet{Murphy2013} for commensal (EMU or WALLABY) surveys using snapshot images (see their Figure~3). 

For the overall 1,476 deg$^2$ sky area, the total number of unique sources is approximately $7\times10^5$. 
We found only a small fraction of sources (0.005 per cent) are variable -- much lower than limit reported by previous searches \citep[e.g.,][]{Mooley2016,Murphy2021}. 
We note that variables or transient sources found in our survey (15-min timescale) are considerably different from objects detected from previous radio transient survey (days to years timescales; e.g., \citealt{Bannister2011,Mooley2016}). 
Our survey is more sensitive to pulsars, flaring radio stars, and enhanced scintillation of extragalactic objects, and not sensitive to extragalactic synchroton transients such as gamma-ray burst and tidal disruption events.  

\begin{figure}
    \includegraphics[width=\columnwidth]{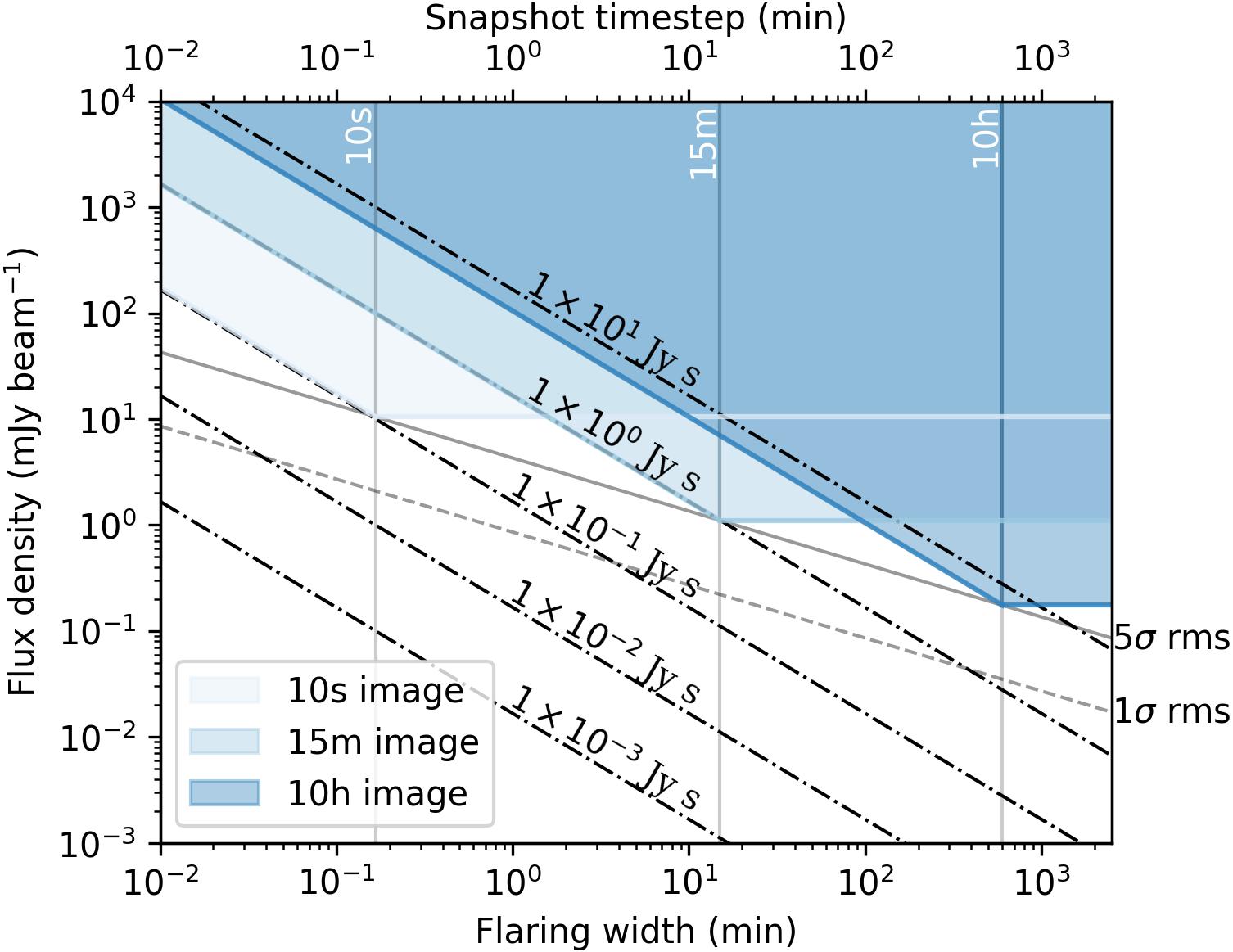}
    \caption{Transient source detection limit of the intra-observation survey strategy. The shallow area represents detectable transient events in different time-length (10\,s, 15\,m, and 10\,h) snapshot images. The dotted-dashed lines represent the fluence of transient events. The grey dashed and solid line represent the $1\sigma$ and $5\sigma$ rms sensitivity for snapshot images in different time-lengths. }
    \label{fig:detection_limit}
\end{figure}

We calculated the detection sensitivity for flaring events with our survey strategy. 
We used the concept of fluence (unit of Jy\,s) in the calculation, defined as the product of the flaring time widths and the peak flux density values. 
As shown in Figure~\ref{fig:detection_limit}, our 15-min snapshots survey are sensitive to transient events with a fluence limit of 1.1\,Jy\,s when the characteristic timescale is shorter than 15 minutes (e.g., for a 100\,Jy burst with a duration of 10\,ms, or a 8\,mJy flare with a duration of 2\,min), or with a peak flux density $\gtrsim$1\,mJy when the characteristic timescale is longer than 15 minutes and shorter than $\sim$10 hours (all are in $5\sigma$ rms threshold). 
For a similar survey using 10-s snapshot images, we are more sensitive to shorter flares, e.g., a fluence limit of 0.18\,Jy\,s for an event with the timescale shorter or comparable to 10 seconds. 
On this timescale we are able to detect fast radio burst-like events \citep[e.g.,][]{Lorimer2007}, or single pulses from ultra-long-period neutron stars (e.g., \citealt{Hurley-Walker2022b,Caleb2022}).

\subsection{Detection rate of stellar objects}

We calculated the detection rate of stellar objects in our survey. 
We found total of eight stars in a 1,476\,deg$^2$ sky area, corresponding to a surface density of $5.4\times10^{-3}$\,deg$^{-2}$.
If we consider a 15-min snapshot equivalent sky coverage (i.e., treat each model-subtracted 15-min image as an independent sky-region snapshot image), we get a surface density of $9.1\times10^{-5}$ \,deg$^{-2}$. 
This rate is an order of magnitude lower than the surface density of $9.66\times10^{-4}$ deg$^{-2}$ reported by \citet{Pritchard2021}, who used 15-min ASKAP images at a similar observing frequency, with all-sky coverage. 
This could be due to multiple factors. 

\begin{description}

\item[{\bf Survey strategy}.]  
\citet{Pritchard2021} conducted a circular polarisation survey (finding any emission in 15-min stokes V images), whereas our survey focused on flux variations within a series of 15-min snapshot images. Our search is less sensitive to the quiescent, polarised emission produced by RS Canum Venaticorum stars, for example.

\item[{\bf Independence of consecutive 15-min snapshots}.] 
Variable radio emission produced either due to rotational modulation of an active region of the stellar corona or stochastic flaring in localised coronal loops is unlikely to be independent on 15-min timescales. 

\item[{\bf Small sampling of the duty cycle distribution}.] 
A search for flaring objects in a deep observation only allows for detection of a small number of stars within the field of view. 
The majority of stars that are detectable by ASKAP have duty cycles well below $0.05$ (Pritchard et al. 2023, in prep.), and therefore repeat sampling of the same field is more likely to sample a small group of low-duty cycle stars compared to a shallow, wide-field search that also covers the less common, highly active stars. 
\end{description}


\subsection{Future plans}
\label{subsec:future_plans}

This work is a demonstration of functionality of our search technique. 
\replace{ASKAP began full survey operations in November 2022}. 
We plan to run our transient detection pipeline commensally on other ASKAP observations, including two wide-field surveys: EMU/POSSUM\footnote{POSSUM is commensal with EMU. } and WALLABY. 
Both of them will cover the entire Southern sky extending as far North as $+30$\degr\ in declination, with 8--10\,h integration time per pointing at 943.5\,MHz or 1367.5\,MHz. 
This will extend our work to an all-sky scale, enabling a comprehensive investigation for minute-timescale transients. 

Future improvements on our pipeline will focus on reducing processing time (An et al., in prep), which is now mainly limited by data transferring time and model generating ($\sim$20 hours). 
We also plan to implement our pipeline into the Pawsey Supercomputing Research Centre's Setonix supercomputer, where data transferring is not needed. 
The standard ASKAP processing for each survey will also be conducted in Setonix, and model images will be generated during this process. 
With these benefits the running time for our transient detection pipeline can be reduced down to a few minutes. 
This will enable rapid trigger and follow-up for individual transient events during full survey operation. 

In this work we identified 11 transients and variables in the $\sim$300\,deg$^2$ EMU pilot observations (i.e., $\sim$1 high variable per EMU field). 
Scaling to the ASKAP full survey, we are likely to identify $\sim$1 source per day assuming one 10-h ASKAP observation per day during full operation, or total of $\sim$1000 highly variable sources in the full EMU survey using the same technique.

We are able to explore different timescales in future, e.g. a 10-s snapshot survey would be able to detect fast radio burst-like events, or ultra-long-period neutron stars. 
Both of these object classes have not been fully understood yet. 
On this timescale it would be possible to identify other unknown transient classes, expanding the understanding of radio time-domain phenomena.

\section{Conclusions} 
\label{sec:conclusions}

We conducted the first radio transients survey to explore minute-timescale variability within hours-long observations. 
We used ASKAP pilot survey data, consisting of 52 8--10\,h observations at around 1\,GHz, with a typical rms level of 30\,\ujy. 
The total sky coverage was 1\,476\,deg$^2$ with overall observing time of 505\,h. 

We detected 38 transients and variable sources: 22 of them (59 per cent) are extreme scintillating objects caused by local plasma screens, eight (21 per cent) are stellar objects, seven (18 per cent) are pulsars, and the remaining one object (2 per cent) is an unidentified transient. 
We observed narrow-band scintles in the dynamic spectra for six (out of seven) pulsars, implying that diffractive scintillation is the main reason for pulsar variations on this timescale. 
The remaining pulsar PSR~J2039$-$5617 is a MSP in a binary system and the variability is due to eclipse. 
For stellar objects, their light-curves have diverse morphologies ranging from short duration bursts (seconds--minutes) to long duration variations ($\gtrsim$hours). 
For 22 variables associated with AGN or galaxies, their variations are likely caused by local plasma screens. 
Their sky distribution is unusual: 21 are clustered in three small regions with an area of 10--50\,deg$^2$ each (overall occupying 5 per cent fraction of our survey coverage). 
The reason for this clustering is unclear. 

Future work on our searching pipeline will focus on the implementation into Setonix supercomputer, which would largely reduce the data processing time down to a few minutes. 
We will also incorporate better methods for dealing with correlated pixels and exploring different timescales as well as detailed rates calculations. 
We expect to identify about 1000 transient and variable sources in the EMU full survey. 
The majority of these are likely to be extreme scintillating objects, which can help to establish an all-sky distribution of local plasma screens. 
We also expect to find $\sim$20 unidentified objects which can possibly be new classes.

\section*{Acknowledgements}
We would like to thank the referee for their helpful comments and suggestions.
We thank Lister Staveley-Smith, Hayley Bignall and Ron Ekers for useful discussions. 
YW is supported by the China Scholarship Council. 
Parts of this research were conducted by the Australian Research Council Centre of Excellence for Gravitational Wave Discovery (OzGrav), project number CE170100004. 

This research was supported by the Sydney Informatics Hub (SIH), a core research facility at the University of Sydney.  
This work used resources of China SKA Regional Centre prototype \citep{An2019,An2022} funded by National SKA Program of China (2022SKA0130103) and the National Key R\&D Programme of China (2018YFA0404603). 
This research has made use of the VizieR catalogue access tool, CDS, Strasbourg, France (DOI : 10.26093/cds/vizier). The original description of the VizieR service was published in \citet{Ochsenbein2000}. 
This research has made use of the SIMBAD database,
operated at CDS, Strasbourg, France \citep{Wenger2000}. 
This research has made use of data obtained from the Chandra Source Catalog, provided by the Chandra X-ray Center (CXC) as part of the Chandra Data Archive.

This scientific work uses data obtained from Inyarrimanha Ilgari Bundara / the Murchison Radio-astronomy Observatory. We acknowledge the Wajarri Yamaji People as the Traditional Owners and native title holders of the Observatory site. CSIRO’s ASKAP radio telescope is part of the Australia Telescope National Facility\footnote{\url{https://ror.org/05qajvd42}}. Operation of ASKAP is funded by the Australian Government with support from the National Collaborative Research Infrastructure Strategy. ASKAP uses the resources of the Pawsey Supercomputing Research Centre. Establishment of ASKAP, Inyarrimanha Ilgari Bundara, the CSIRO Murchison Radio-astronomy Observatory and the Pawsey Supercomputing Research Centre are initiatives of the Australian Government, with support from the Government of Western Australia and the Science and Industry Endowment Fund.
This paper includes archived data obtained through the CSIRO ASKAP Science Data Archive, CASDA (http://data.csiro.au). 

This research has made use of 
\textsc{aplpy}~\citep{Robitaille2012}, 
\textsc{astropy}~\citep{AstropyCollaboration2013,AstropyCollaboration2018}, 
\textsc{astroquery}~\citep{Ginsburg2019}, 
\textsc{matplotlib}~\citep{Hunter2007},
\textsc{mocpy}\footnote{\url{https://github.com/cds-astro/mocpy}}, 
\textsc{numpy}~\citep{Harris2020}, 
\textsc{pygsm}~\citep{Price2016}, 
and \textsc{scipy}~\citep{2020SciPy-NMeth}.

\section*{Data Availability}

All of the ASKAP data used in this paper can be accessed through the CSIRO ASKAP Science Data Archive (CASDA\footnote{\url{https://data.csiro.au/dap/public/casda/casdaSearch.zul}}), using the project codes and SBIDs listed in Table~\ref{tab:observations}.



\bibliographystyle{mnras}
\bibliography{example} 








\bsp	
\label{lastpage}
\end{document}

%% file: table_observations.tex
\begin{table*}
    \centering
	\caption{Details of the 52 ASKAP pilot observations we used in this work, including the scheduling block ID (SBID), field name, coordinates of field center in J2000, central frequency (unit of MHz), observing start time in UTC, and the duration of each observation. The SBID and project code can be used to access the data via the CSIRO ASKAP Science Data Archive.}
	\label{tab:observations}
    \begin{tabular}{cccccccc}
        \hline
        SBID & Field Name & RA  & DEC  & Cent. Freq.  & Star Time (UTC) & Duration  & Project Code \\
         & & (hh:mm:ss) & (dd:mm:ss) & (MHz) & & (hh:mm:ss) \\
        \hline
        9287 & EMU\_2059-51 & 21:00:30.210 & $-$51:11:32.074 & 943.5 & 2019-07-15 13:49:08 & 10:00:11 & AS101 \\
        9325 & EMU\_2034-60 & 20:35:13.283 & $-$60:24:59.883 & 943.5 & 2019-07-17 13:23:37 & 10:00:11 & AS101 \\
        9351 & EMU\_2042-55 & 20:42:40.597 & $-$55:48:29.838 & 943.5 & 2019-07-18 13:35:59 & 10:00:11 & AS101 \\
        9410 & EMU\_2115-60 & 21:16:21.856 & $-$60:24:59.883 & 943.5 & 2019-07-24 12:23:45 & 10:00:11 & AS101 \\
        9434 & EMU\_2132-51 & 21:33:13.847 & $-$51:11:32.074 & 943.5 & 2019-07-25 13:16:38 & 10:00:11 & AS101 \\
        9437 & EMU\_2027-51 & 20:27:46.574 & $-$51:11:32.074 & 943.5 & 2019-07-26 11:26:23 & 10:00:11 & AS101 \\
        9442 & EMU\_2118-55 & 21:18:40.597 & $-$55:48:29.838 & 943.5 & 2019-07-27 12:13:06 & 10:00:11 & AS101 \\
        9501 & EMU\_2156-60 & 21:57:30.427 & $-$60:24:59.883 & 943.5 & 2019-08-02 12:28:01 & 10:00:01 & AS101 \\
        10083 & EMU\_2154-55 & 21:54:40.597 & $-$55:48:29.838 & 943.5 & 2019-10-03 08:21:36 & 10:00:11 & AS101 \\
        10635 & EMU\_2205-51 & 22:05:57.482 & $-$51:11:32.074 & 943.5 & 2019-11-24 06:10:19 & 10:00:11 & AS101 \\
        10636 & Abell\_3266 & 04:32:25.107 & $-$61:32:30.178 & 943.5 & 2019-11-24 16:11:40 & 6:45:05 & AS101 \\
        10269 & Hydra\_1B & 10:17:51.611 & $-$27:49:21.388 & 1367.5 & 2019-10-25 19:23:48 & 8:00:04 & AS102 \\
        10609 & Hydra\_1A & 10:15:48.413 & $-$27:22:48.299 & 1367.5 & 2019-11-20 18:33:48 & 8:00:14 & AS102 \\
        10612 & Hydra\_2A & 10:39:24.807 & $-$27:22:48.299 & 1367.5 & 2019-11-21 18:52:37 & 8:00:04 & AS102 \\
        10626 & Hydra\_2B & 10:41:28.005 & $-$27:49:21.388 & 1367.5 & 2019-11-23 18:47:26 & 8:00:04 & AS102 \\
        10736 & NGC4636\_2A & 12:38:02.392 & +04:57:09.446 & 1367.5 & 2019-12-05 19:56:50 & 8:37:24 & AS102 \\
        10809 & NGC4636\_1A & 12:38:02.134 & $-$00:26:53.976 & 1367.5 & 2019-12-12 19:28:22 & 8:39:03 & AS102 \\
        10812 & NGC4636\_1B & 12:39:50.199 & $-$00:53:55.037 & 1367.5 & 2019-12-13 19:26:17 & 8:38:53 & AS102 \\
        11816 & Norma\_1A & 16:16:35.826 & $-$59:29:14.718 & 1367.5 & 2020-02-14 18:56:01 & 8:00:04 & AS102 \\
        11832 & Norma\_1B & 16:20:16.779 & $-$59:54:55.768 & 1367.5 & 2020-02-15 18:54:46 & 8:00:14 & AS102 \\
        12193 & Norma\_2A & 16:55:30.959 & $-$59:29:14.718 & 1367.5 & 2020-03-13 17:43:41 & 8:00:04 & AS102 \\
        12209 & Norma\_2B & 16:59:11.914 & $-$59:54:55.768 & 1367.5 & 2020-03-14 17:43:36 & 8:00:04 & AS102 \\  
        9945 & POSSUM\_pilot\_2032-54 & 20:32:37.797 & $-$54:37:17.255 & 1367.5 & 2019-09-19 10:03:21 & 9:34:08 & AS103 \\
        9962 & POSSUM\_2045-50 & 20:46:04.907 & $-$50:44:45.633 & 1367.5 & 2019-09-20 09:05:46 & 10:00:01 & AS103 \\
        9972 & POSSUM\_2038-58 & 20:39:00.327 & $-$58:26:20.890 & 1367.5 & 2019-09-21 09:12:02 & 10:00:11 & AS103 \\
        9975 & POSSUM\_2057-54 & 20:57:37.797 & $-$54:37:17.255 & 1367.5 & 2019-09-22 09:17:35 & 10:00:01 & AS103 \\
        9983 & POSSUM\_2113-50 & 21:13:25.953 & $-$50:47:31.669 & 1367.5 & 2019-09-23 09:21:06 & 10:00:11 & AS103 \\
        9992 & POSSUM\_2110-58 & 21:11:16.872 & $-$58:29:41.342 & 1367.5 & 2019-09-24 09:32:25 & 10:00:11 & AS103 \\
        10007 & POSSUM\_2126-54 & 21:27:15.485 & $-$54:40:10.601 & 1367.5 & 2019-09-25 09:35:22 & 10:00:11 & AS103 \\
        10040 & POSSUM\_2156-54 & 21:56:55.161 & $-$54:42:41.796 & 1367.5 & 2019-09-28 09:53:15 & 10:00:11 & AS103 \\
        10043 & POSSUM\_2207-50 & 22:08:12.569 & $-$50:52:12.774 & 1367.5 & 2019-09-29 09:52:20 & 10:00:01 & AS103 \\
        10168 & POSSUM\_2140-50 & 21:40:48.566 & $-$50:50:01.774 & 1367.5 & 2019-10-16 08:18:08 & 8:43:22 & AS103 \\
        9596 & LIGO\_0556-3309 & 05:56:28.987 & $-$33:11:53.141 & 943.5 & 2019-08-15 19:35:14 & 10:00:01 & AS111 \\
        9602 & S190814bv & 00:50:43.726 & $-$25:19:04.988 & 943.5 & 2019-08-16 14:11:23 & 10:38:10 & AS111 \\
        9649 & S190814bv & 00:50:43.726 & $-$25:19:04.988 & 943.5 & 2019-08-23 13:43:55 & 10:37:40 & AS111 \\
        9910 & S190814bv & 00:50:43.726 & $-$25:19:04.988 & 943.5 & 2019-09-16 12:09:33 & 10:37:20 & AS111 \\
        10463 & S190814bv & 00:50:43.726 & $-$25:19:04.988 & 943.5 & 2019-11-07 08:45:10 & 10:37:20 & AS111 \\
        12704 & S190814bv & 00:50:43.726 & $-$25:19:04.988 & 943.5 & 2020-04-03 23:00:00 & 34:32:46$^a$ & AS111 \\
        13570 & S190814bv & 00:58:07.108 & $-$23:44:33.168 & 943.5 & 2020-04-29 21:41:11 & 10:00:11 & AS107$^b$ \\
        15191 & S190814bv & 00:50:43.726 & $-$25:19:04.988 & 943.5 & 2020-07-03 17:01:21 & 10:30:12 & AS111 \\
        18912 & S190814bv & 00:50:43.726 & $-$25:19:04.988 & 943.5 & 2020-11-29 07:15:32 & 10:52:26$^a$ & AS111 \\
        18925 & S190814bv & 00:50:43.726 & $-$25:19:04.988 & 943.5 & 2020-11-28 09:18:30 & 8:53:19 & AS111 \\
        27379 & S190814bv & 00:50:37.426 & $-$25:17:00.371 & 943.5 & 2021-05-29 19:23:45 & 10:30:12 & AS111 \\
        10108 & eFEDS\_low\_T1-1A & 09:03:59.506 & +04:39:12.756 & 887.5 & 2019-10-05 20:09:37 & 8:38:53 & AS112 \\
        10123 & eFEDS\_low\_T0-1A & 08:38:44.323 & +04:38:40.490 & 887.5 & 2019-10-07 19:47:17 & 8:41:23 & AS112 \\
        10126 & eFEDS\_low\_T0-0A & 08:38:49.891 & $-$01:39:23.893 & 887.5 & 2019-10-08 19:36:54 & 8:38:53 & AS112 \\
        10129 & eFEDS\_low\_T1-0A & 09:03:59.635 & $-$01:38:52.067 & 887.5 & 2019-10-09 20:24:35 & 7:32:22 & AS112 \\
        10132 & eFEDS\_low\_T2-1A & 09:29:15.225 & +04:38:28.707 & 887.5 & 2019-10-10 20:27:09 & 8:39:03 & AS112 \\
        10135 & eFEDS\_low\_T2-0A & 09:29:09.378 & $-$01:39:24.359 & 887.5 & 2019-10-11 20:29:23 & 8:33:45 & AS112 \\
        10137 & eFEDS\_low\_T1-0A & 09:03:59.635 & $-$01:38:52.067 & 887.5 & 2019-10-12 19:51:28 & 8:39:03 & AS112 \\
        25035 & TESS\_Sector-36A & 04:24:00.754 & $-$70:02:20.506 & 887.5 & 2021-03-20 03:05:00 & 13:00:00 & AS113 \\
        25077 & TESS\_Sector-36B & 05:08:01.681 & $-$60:01:10.217 & 887.5 & 2021-03-21 03:31:14 & 12:57:11 & AS113 \\
        \hline
        \multicolumn{8}{p{0.9\textwidth}}{$^a$ These datasets consist of two observations with $\gtrsim$ hours-long gap between them. The overall integration time is 15.3\,h for SB12704 and 7.1\,h for SB18912. See details in \citet{Dobie2022}}\\
        \multicolumn{8}{p{0.9\textwidth}}{\footnotesize$^b$ This is a test observation to rule out potential instrumental effects, as described in \citet{Wang2021b}. This observation is under a different project code AS107 (the ASKAP Variables and Slow Transients; VAST; \citealt{Murphy2013})}\\
    \end{tabular}
\end{table*}

%% file: table_candidates.tex
\begin{landscape}
\begin{table}
\centering
    \caption{Transients and variables identified in our survey. 
    We list their coordinates, variability metrics $\eta$ and $m$, and flux density $S_\textrm{deep}$ (as described in Section~\ref{subsec:lightcurve_analysis}). 
    We list all of observations that covered the source location, and mark observations where we detected no variable behaviour as *. 
    The listed beam has the beam centre closest to the source location. 
    We provide the SIMBAD ID, the \textit{WISE} cross-ID, or the LS object ID from the DESI Legacy Imaging Survey DR8 catalogue \citep{Dey2019}. We list the mean and standard deviation of the photometric redshift distributions when available (from \citealt{Duncan2022}). }
    \label{tab:candidates}
    \begin{tabular}{lccccclll}
        \hline
        Name & RA (J2000) & DEC (J2000) & $\eta$ & $m$ & $S_\mathrm{deep}$ & SBID (beam number) & Identification & Notes \\
         & (hh:mm:ss) & (dd:mm:ss) & & & (mJy\,beam$^{-1}$) & & & \\
        \hline
        Pulsars &  &  &  &  &  &  &  &  \\
        VAST~J083705.6$+$061015 & 08:37:05.68 & $+$06:10:15.63 & 106.43 & 0.34 & $7.19\pm0.06$ & 10123 (B5) & PSR~J0837$+$0610 & Nulling pulsar \\
        VAST~J092214.0$+$063824 & 09:22:14.03 & $+$06:38:24.17 & 115.44 & 0.22 & $10.86\pm0.06$ & 10132 (B4) & PSR~J0922$+$0638 & - \\
        VAST~J170416.8$-$601934 & 17:04:16.85 & $-$60:19:34.79 & 26.03 & 0.26 & $2.58\pm0.03$ & 12193 (B10); 12209 (B3) & PSR~J1704$-$6016 & - \\
        VAST~J203934.8$-$561710 & 20:39:34.86 & $-$56:17:10.37 & 12.96 & 0.90 & $0.28\pm0.03$ & 9351 (B14); 9945 (B4)* & PSR~J2039$-$5617 & Eclipsing MSP \\
        VAST~J212922.8$-$572114 & 21:29:22.80 & $-$57:21:14.41 & 2.96 & 0.14 & $1.45\pm0.05$ & 9442 (B4) & PSR~J2129$-$5721 & MSP \\
        VAST~J213657.5$-$504656 & 21:36:57.58 & $-$50:46:56.83 & 4.40 & 0.13 & $1.34\pm0.03$ & 9434 (B21); 10168 (B14) & PSR~J2136$-$5046 & - \\
        VAST~J214435.6$-$523707 & 21:44:35.67 & $-$52:37:07.73 & 12.91 & 0.22 & $1.53\pm0.03$ & 9434 (B10); 10040 (B30)*; 10168 (B4) & PSR~J2144$-$5237 & MSP \\
        \hline
        Stars &  &  &  &  &  &  &  &  \\
        VAST~J005830.0$-$275132 & 00:58:30.03 & $-$27:51:32.65 & 5.91 & - & - & 12704 (B5)$^a$ & CD-28~302 & dM4 \\
        VAST~J044649.5$-$603408 & 04:46:49.50 & $-$60:34:08.13 & 2.32 & 1.06 & $0.26\pm0.04$ & 25077 (B32); 10636 (B29)* & BPS~CS~29520-0077$^b$ & M1.5V \\
        VAST~J055306.7$-$334803 & 05:53:06.78 & $-$33:48:03.50 & 5.04 & 0.65 & $0.23\pm0.03$ & 9596 (B14) & \replace{$\lambda$ Columbae} & B5V \\
        VAST~J104918.8$-$250924 & 10:49:18.83 & $-$25:09:24.51 & 2.59 & - & - & 10612 (B21) & 2MASS~J10491880$-$2509235 & M4.9 \\
        VAST~J170946.3$-$595740 & 17:09:46.30 & $-$59:57:40.26 & 2.48 & 0.17 & $0.88\pm0.03$ & 12209 (B9); 12193 (B9)* & UPM~J1709$-$5957 & -- \\
        VAST~J202920.3$-$512946 & 20:29:20.30 & $-$51:29:46.37 & 3.05 & 0.39 & $0.30\pm0.04$ & 9437 (B15) & WT~713 & M dwarf \\
        VAST~J213835.4$-$505111 & 21:38:35.49 & $-$50:51:11.64 & 5.41 & 0.53 & $0.31\pm0.02$ & 10168 (B14); 9434 (B21)* & RX~J2138.5$-$5050 & M7/8 candidate \\
        VAST~J215741.1$-$510030 & 21:57:41.12 & $-$51:00:30.72 & 3.62 & 0.28 & $0.58\pm0.03$ & 10635 (B19); 10043 (B13)* & CD-51~13128 & M dwarf binary \\
        \hline
        AGN/Galaxies &  &  &  &  &  &  &  &  \\
        VAST~J004711.9$-$262350 & 00:47:11.95 & $-$26:23:50.62 & 22.97 & 0.06 & $9.69\pm0.03$ & S190814bv fields (B8) & WISEA~J004712.05$-$262350.5 & Quasar ($z=1.11\pm0.47$) \\
        VAST~J004929.7$-$241846 & 00:49:29.79 & $-$24:18:46.67 & 16.09 & 0.05 & $10.34\pm0.04$ & S190814bv fields (B26) & 3HSP~J004929.9$-$241844 & Blazar ($z=0.78\pm0.50$) \\
        VAST~J005716.8$-$251424 & 00:57:16.83 & $-$25:14:24.47 & 16.93 & 0.05 & $12.98\pm0.04$ & S190814bv fields (B22) & WISEA~J005716.92$-$251424.4 & See \citet{Wang2021b} \\
        VAST~J005800.8$-$235449 & 00:58:00.87 & $-$23:54:49.03 & 162.47 & 0.25 & $8.80\pm0.04$ & S190814bv fields (B29) & WISEA~J005800.99$-$4235448.0 & See \citet{Wang2021b} \\
        VAST~J005806.6$-$234744 & 00:58:06.69 & $-$23:47:44.47 & 59.42 & 0.15 & $8.40\pm0.04$ & S190814bv fields (B29) & LS~DR8~8000197415002484 & See \citet{Wang2021b} \\
        VAST~J005808.9$-$233453 & 00:58:08.93 & $-$23:34:53.77 & 6.22 & 0.18 & $1.47\pm0.04$ & S190814bv fields (B34) & WISEA~J005808.68$-$233453.0 & See \citet{Wang2021b} \\
        VAST~J005811.9$-$233735 & 00:58:11.96 & $-$23:37:35.69 & 176.56 & 0.41 & $4.89\pm0.03$ & S190814bv fields (B34) & WISEA~J005812.03$-$233735.6 & See \citet{Wang2021b} \\
        VAST~J042535.9$-$590643 & 04:25:35.96 & $-$59:06:43.85 & 45.41 & 0.06 & $20.42\pm0.03$ & 10636 (B31) & WISEA~J042535.97$-$590642.4 & Quasar \\
        VAST~J042932.8$-$593118 & 04:29:32.89 & $-$59:31:18.61 & 5.35 & 0.06 & $4.98\pm0.05$ & 10636 (B32) & LS-DR8~8000045911001631 & Quasar ($z=2.42\pm0.36$) \\
        VAST~J043628.3$-$595154 & 04:36:28.34 & $-$59:51:54.25 & 13.47 & 0.06 & $8.66\pm0.04$ & 10636 (B33) & 2MASX~J04362838$-$5951531 & Galaxy ($z=0.058\pm0.021$) \\
        VAST~J045152.3$-$590015 & 04:51:52.31 & $-$59:00:15.22 & 5.34 & 0.06 & $5.75\pm0.03$ & 25077 (B34); 10636 (B35)* & WISEA~J045152.41$-$590014.8 & Quasar ($z=1.56\pm0.39$) \\
        VAST~J045922.0$-$563419 & 04:59:22.05 & $-$56:34:19.04 & 5.91 & 0.07 & $6.52\pm0.05$ & 25077 (B16) & - & - \\
        VAST~J050316.8$-$585326 & 05:03:16.81 & $-$58:53:26.42 & 5.25 & 0.06 & $5.57\pm0.04$ & 25077 (B0) & WISEA~J050316.89$-$585325.1 & Galaxy ($z=0.85\pm0.09$) \\
        VAST~J050450.0$-$585328 & 05:04:50.04 & $-$58:53:28.51 & 13.03 & 0.13 & $3.31\pm0.03$ & 25077 (B0) & WISEA~J050450.20-585328.2 & Galaxy ($z=0.17\pm0.02$) \\
        VAST~J051541.5$-$605936 & 05:15:41.52 & $-$60:59:36.16 & 5.68 & 0.06 & $4.74\pm0.03$ & 25077 (B9) & WISEA~J051541.65$-$605935.9 & Quasar ($z=1.71\pm0.47$) \\
        VAST~J052646.0$-$595536 & 05:26:46.07 & $-$59:55:36.05 & 5.37 & 0.11 & $2.22\pm0.03$ & 25077 (B23) & WISEA~J052646.15$-$595535.7 & Galaxy ($z=0.29\pm0.02$) \\
        VAST~J052845.8$-$592003 & 05:28:45.86 & $-$59:20:03.60 & 19.15 & 0.05 & $11.11\pm0.03$ & 25077 (B7) & 4FGL~J0528.7$-$5920 & Blazar \\
        VAST~J053003.0$-$592700 & 05:30:03.01 & $-$59:27:00.75 & 3.34 & 0.10 & $1.99\pm0.03$ & 25077 (B22) & WISEA~J053003.05$-$592701.2 & Galaxy ($z=1.13\pm0.36$) \\
        VAST~J104933.6$-$245653 & 10:49:33.63 & $-$24:56:53.30 & 38.84 & 0.10 & $12.37\pm0.04$ & 10612 (B21) & WISEA~J104933.63$-$245651.8 & Quasar \\
        VAST~J214659.5$-$485325 & 21:46:59.54 & $-$48:53:25.94 & 9.69 & 0.05 & $10.81\pm0.04$ & 9434 (B35); 10168 (B34)* & WISEA~J214659.48$-$485325.4 & Galaxy ($z=1.01\pm0.67$) \\
        VAST~J220056.5$-$500545 & 22:00:56.52 & $-$50:05:45.53 & 76.02 & 0.06 & $24.25\pm0.08$ & 10635 (B26); 10043 (B25)* & LS-DR8~8000078211000459 & Quasar ($z=1.27\pm0.47$) \\
        VAST~J221429.1$-$494845 & 22:14:29.13 & $-$49:48:45.60 & 4.21 & 0.10 & $2.44\pm0.03$ & 10635 (B29); 10043 (B28)* & WISEA~J221429.13$-$494845.1 & Quasar \\
        \hline
        Unidentified &  &  &  &  &  &  &  &  \\
        VAST~J205139.0$-$573906 & 20:51:39.00 & $-$57:39:06.38 & 3.32 & 0.32 & $0.50\pm0.03$ & 9972 (B29); 9325 (B25)*; 9351 (B4)* & - & ? \\
        \hline
        \multicolumn{9}{p{1.3\textwidth}}{$^a$ No detection in the time-averaged model image. Only detected one single flare in SB12704 although covered by other S190814bv fields.   \footnotesize$^b$ Detected in the original paper by \citet{Rigney2022}. }
    \end{tabular}
\end{table}
\end{landscape}

%% file: figure_lightcurves.tex
\begin{figure*}
    \includegraphics[width=\textwidth]{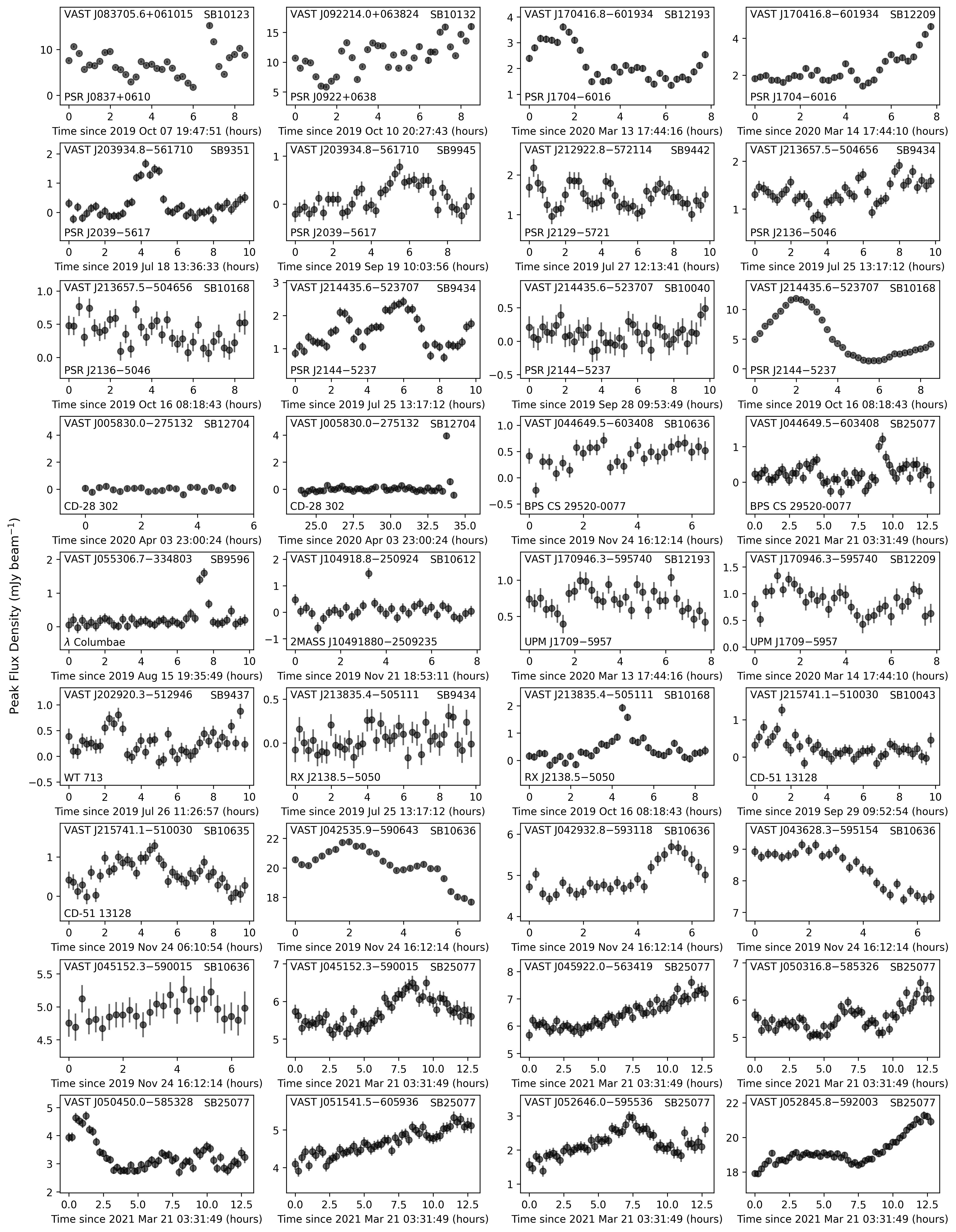}
    \caption{Light-curves of variables and transient objects identified in our survey. The start time shown in x axis label is the UTC time. These light-curves follows the order of objects listed in Table~\ref{tab:candidates}, but we moved sources in S190814bv fields to the end as they normally have $\sim$10 observations. }
    \label{fig:lightcurves}
\end{figure*}

\begin{figure*}\ContinuedFloat
    \includegraphics[width=\textwidth]{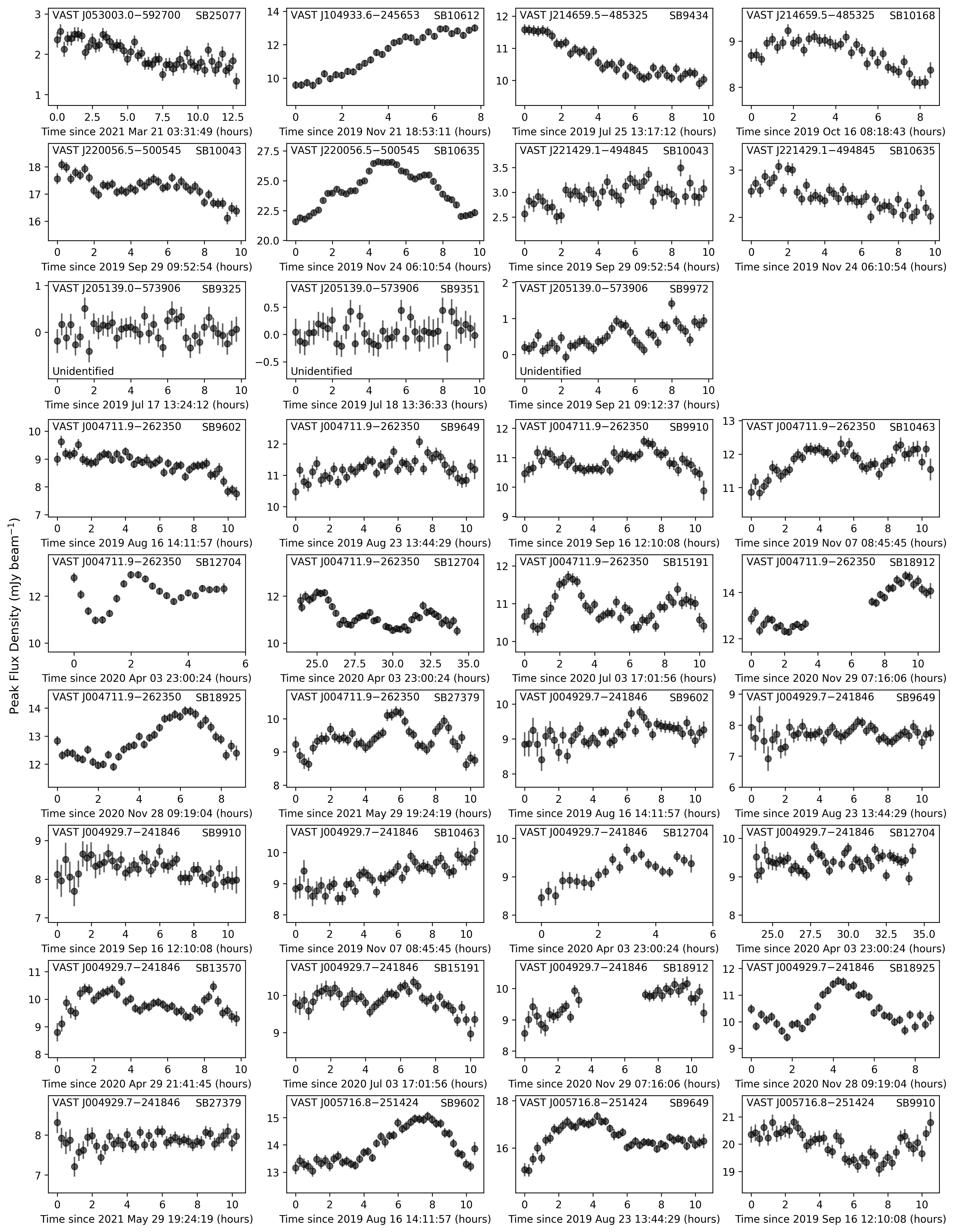}
    \caption{(continued)}
\end{figure*}

\begin{figure*}\ContinuedFloat
    \includegraphics[width=\textwidth]{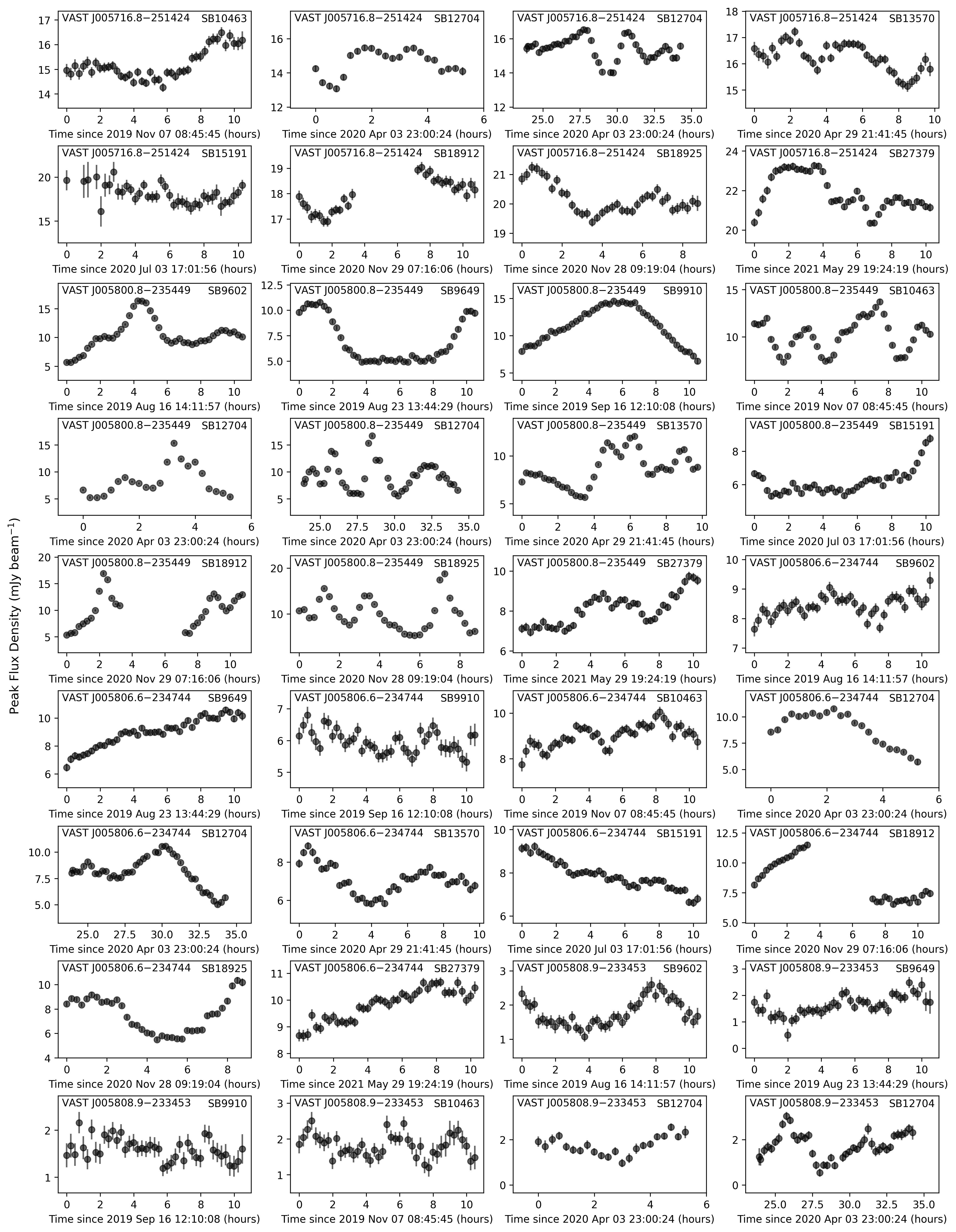}
    \caption{(continued)}
\end{figure*}

\begin{figure*}\ContinuedFloat
    \includegraphics[width=\textwidth]{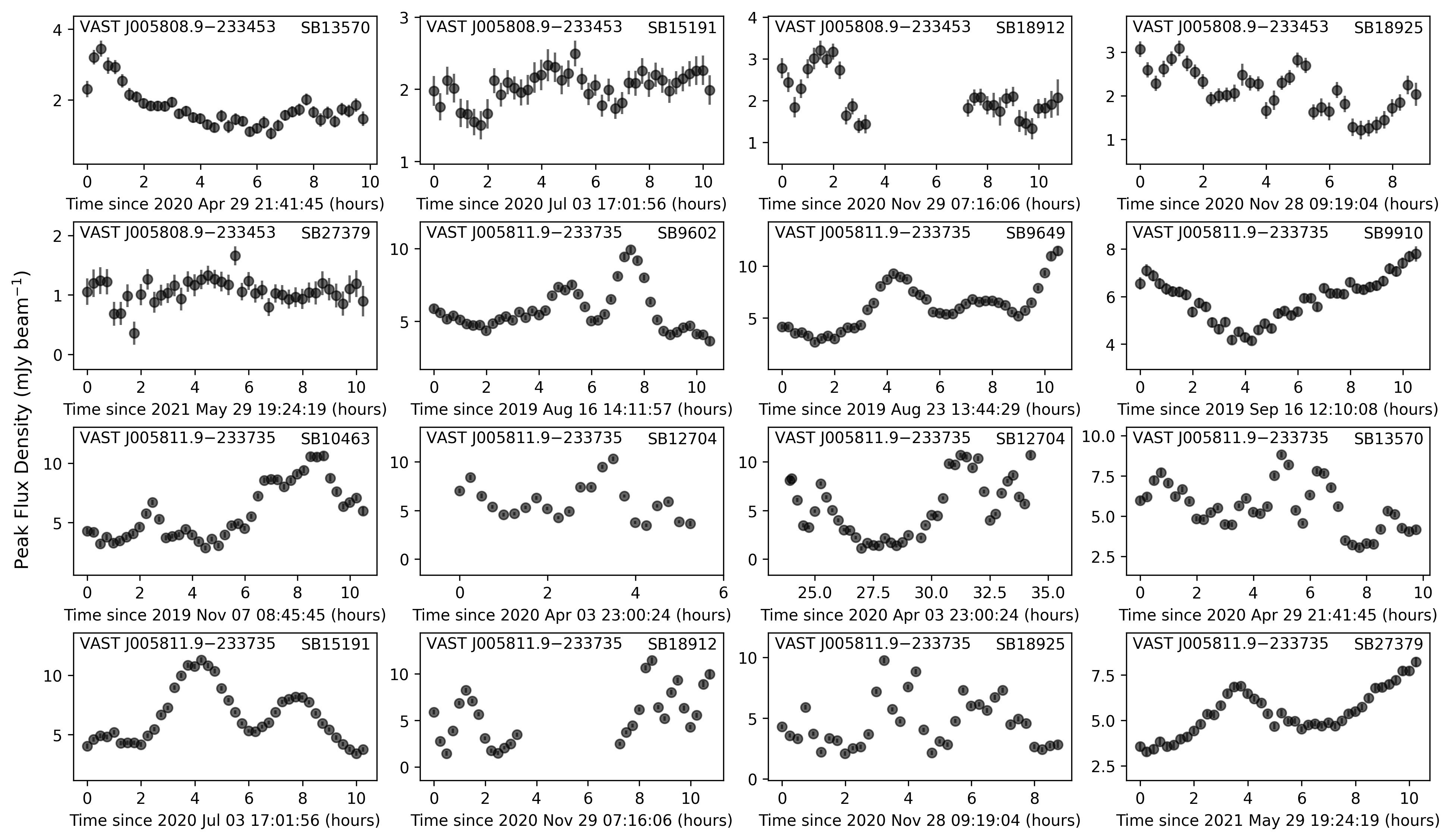}
    \caption{(continued)}
\end{figure*}

%% file: table_stars.tex
\begin{table}
    \centering
    \caption{\replace{Table of detected radio stars. We list the estimated radio luminosities and variability timescales. }}
    \label{tab:stars}
    \begin{tabular}{lccccccc}
        \hline
        Name & Radio Luminosity & Timescale \\
         & (erg\,s$^{-1}$\,Hz$^{-1}$) \\
        \hline 
        CD-28~302 & $7.6\times10^{15}$ & 30\,s; 4\,min \\
        BPS~CS~29520-0077 & $8.8\times10^{14}$ & $\sim$1\,h \\
        $\lambda$ Columbae & $2.1\times10^{16}$ & $\sim$1\,h \\
        2MASS J10491880$-$2509235 & $4.6\times10^{17}$ & 1\,min \\
        UPM J1709$-$5957 & $3.2\times10^{14}$ & a few hours \\
        WT 713 & $1.5\times10^{15}$ & $\sim$1\,h \\
        RX J2138.5$-$5050 & $4.7\times10^{15}$ & $\sim$1\,h \\
        CD-51 13128 & $3.4\times10^{14}$ & a few hours \\
        \hline
    \end{tabular}
\end{table}